\documentclass[%
superscriptaddress,
preprint,
 amsmath,amssymb,
 aps,
prb,
]{revtex4-1}

\usepackage{graphicx}
\usepackage{dcolumn}
\usepackage{bm}
\usepackage{xcolor}

\begin{document}

\preprint{ }

\title{Hanbury Brown and Twiss Exchange Correlations in Graphene Box}

\author{Teemu Elo}
\email{teemu.elo@aalto.fi}
\author{Zhenbing Tan}%
\thanks{Present address: Shenzhen Institute for Quantum Science and Engineering, and Department of Physics, Southern University of Science and Technology, Shenzhen 518055, China.}
\affiliation{Low Temperature Laboratory, Department of Applied Physics, Aalto University, Espoo, Finland}
\author{Ciprian Padurariu}
\affiliation{Low Temperature Laboratory, Department of Applied Physics, Aalto University, Espoo, Finland}
\affiliation{Institute for Complex Quantum Systems and IQST, Ulm University, Ulm, Germany}
\author{Fabian Duerr}
\affiliation{Physikalisches Institut (EP3), University of W{\"u}rzburg, W{\"u}rzburg, Germany}%
\author{Dmitry S. Golubev}
\affiliation{Low Temperature Laboratory, Department of Applied Physics, Aalto University, Espoo, Finland}
\author{Gordey B. Lesovik}
\affiliation{Moscow Institute of Physics and Technology, Moscow 141700,
Russian Federation}
\author{Pertti Hakonen}
\email{pertti.hakonen@aalto.fi}
\affiliation{Low Temperature Laboratory, Department of Applied Physics, Aalto University, Espoo, Finland}


\begin{abstract}
Quadratic detection in linear mesoscopic transport systems produces cross terms that can be viewed as interference signals reflecting statistical properties of charge carriers. In electronic systems these cross term interferences arise from exchange effects due to Pauli principle. Here we demonstrate fermionic Hanbury Brown and Twiss (HBT) exchange phenomena due to indistinguishability of charge carriers in a diffusive graphene system. These exchange effects are verified using current-current cross correlations in combination with regular shot noise (autocorrelation) experiments at microwave frequencies. Our results can be modeled using semiclassical analysis for a square-shaped metallic diffusive conductor, including contributions from contact transparency. The experimentally determined HBT exchange factor values lie between the calculated ones for coherent and hot electron transport.

\end{abstract}

\maketitle


\section{Introduction}

Shot noise is a widely used characterization method in nanophysics, as it can provide more information of the charge transport than conventional conductance or thermal noise measurements  \cite{Kogan1996,Blanter2000,Martin2005,Lesovik2011}. Multiterminal current-current correlation experiments provide additional insight to intrinsic characteristics of charge carriers in mesoscopic systems. For example, they allow one to distinguish bosonic and fermionic carriers \cite{Liu1998,Oliver1999,Henny1999}. 

Many of the noise and cross correlation experiments probing fundamental properties of the charge carriers have been performed using edge states in the quantum Hall regime, in which quantum point contacts with tunable transparency control the propagation of coherent beams of electrons or composite fermions \cite{Glattli2005}. In this setup, one can perform two-particle scattering experiments and observe Hanbury Brown and Twiss \cite{R.HanburyBrown1956} (HBT) interference effects in current-current cross correlation \cite{Neder2007}, which are not visible in Aharonov-Bohm conductance experiments. In a regular mesoscopic conductor the phase-dependent phenomena in two-particle scattering events are averaged out over many possible trajectories\cite{Blanter1997}. However, even after such averaging current-current cross correlations in different terminals are affected by Fermi statistics of electrons in a non-trivial way. One well known consequence of Fermi statistics is the negative sign of cross correlations between the currents in different terminals \cite{Blanter2000}. In this work we investigate another interesting consequence -- non-additive nature of cross correlations \cite{Blanter1997,Sukhorukov1999} -- in a HBT setup \cite{R.HanburyBrown1956,Neder2007} with two sources and two detectors attached to a diffusive graphene flake. Below we will refer to the non-additivity of the noise cross correlations as HBT exchange effect.

To our knowledge, only one experiment has so far addressed HBT exchange effects in diffusive conductors. Cross correlations and HBT exchange were measured in a cross-shaped graphene conductor in which the charge carrier density, and thereby the screening of impurities, could be tuned by the back gate voltage \cite{Tan2018}. According to the theory, in a diffusive conductor with cross geometry the paths of scattering electrons are quite restricted, and the HBT exchange effect should disappear \cite{Blanter1997,Sukhorukov1999}. However, the experiment showed a finite exchange effect, which was attributed to an appreciable mean free path of electrons, comparable to the size of the crossing.

In charge neutral graphene, ideally, electrical transport takes place via evanescent waves, the distribution of which mimics diffusive electron transport \cite{Katsnelson2006,Tworzydo2006,San-Jose2007,Lewenkopf2008}. Since the evanescent waves may propagate to both measuring terminals, special cross correlations are obtained in graphene near the charge neutrality point (CNP) \cite{Laakso2008}. According to the tight binding calculations of Ref.~\cite{Laakso2008}, there is negative HBT exchange effect at the Dirac point. 
Instead of diffusive-like shot noise due to evanescent waves, experiments have shown more complex behavior in graphene \cite{DiCarlo2008,Danneau2008,Danneau2009}. For graphene ribbons, Coulomb blockade effects and localization have been found to influence the shot noise results substantially \cite{Danneau2010}. Therefore, also shot noise cross-correlations can be expected to differ from those appearing in pure diffusive transport and to exhibit features inherent to disordered graphene samples.

In this work we study HBT exchange effect in coherent square-shaped graphene conductor with short mean free path and diffusive transport of electrons.
We measure both current-current cross correlations at microwave frequencies and regular shot noise of the contacts (autocorrelation). We model our results using semiclassical analysis for a  diffusive coherent conductor, in which the noise arises locally due to the non-equilibrium distribution of electrons. We repeat the analysis in the hot electron regime, where the noise is characterized by local temperature distribution. Best agreement between experiment and theory is obtained in crossover regime between the coherent and hot electron models.

This article is organized as follows. 
We start with theoretical background (Sec.~II), and outline the basics of shot noise, cross correlations, and the Hanbury Brown and Twiss exchange effect in fermionic systems. In Sec. II A, we describe briefly semiclassical analysis and present our models for coherent and hot electron regimes. The parameters for the numerical noise calculations are obtained from conductance distribution of our sample, analyzed in Sec.~II B, while the noise calculations are presented in Sec.~II C. 
Our experimental methods are concisely covered in Sec.~III, while results are presented in Sec. IV. The discussion in Sec.~V includes connections of our work to other noise experiments and discusses a few theoretical issues relevant for the bias and gate voltage dependence of our data. Sec.~VI concludes the paper.

\section{Theoretical background}

Random flow of electrons with charge $e$ can be described as an uncorrelated
Poisson process \cite{Schottky1918}, which gives rise to  the spectral density of the shot noise,
$S_I=2eI$, where $I$ is the current through the conductor. In contrast to
thermal fluctuations in mesoscopic conductors, shot noise provides information on the basic transport properties beyond
the linear response theory coefficients such as conductance.
In mesoscopic systems, shot noise can become sub-Poissonian under the influence of interactions or correlations, for example,
imposed by the Pauli principle \cite{Khlus1987, Landauer1989, Lesovik1989, Yurke1990, Buttiker1990}. The ensuing noise spectral density can be written as $S_I= F2eI$ where $F$ denotes the so called Fano factor.  In a tunnel junction with low transmission, $F=1$ because  electron tunneling in such a junction is a Poissonian process \cite{Blanter2000}.
In ballistic conductor the shot noise is fully suppressed, while suppression down to $F=1/3$ is found in diffusive conductors  \cite{Beenakker1992,Nagaev1992,Blanter2000}.

\begin{figure}
  \begin{center}
    \leavevmode
\includegraphics[width=\linewidth]{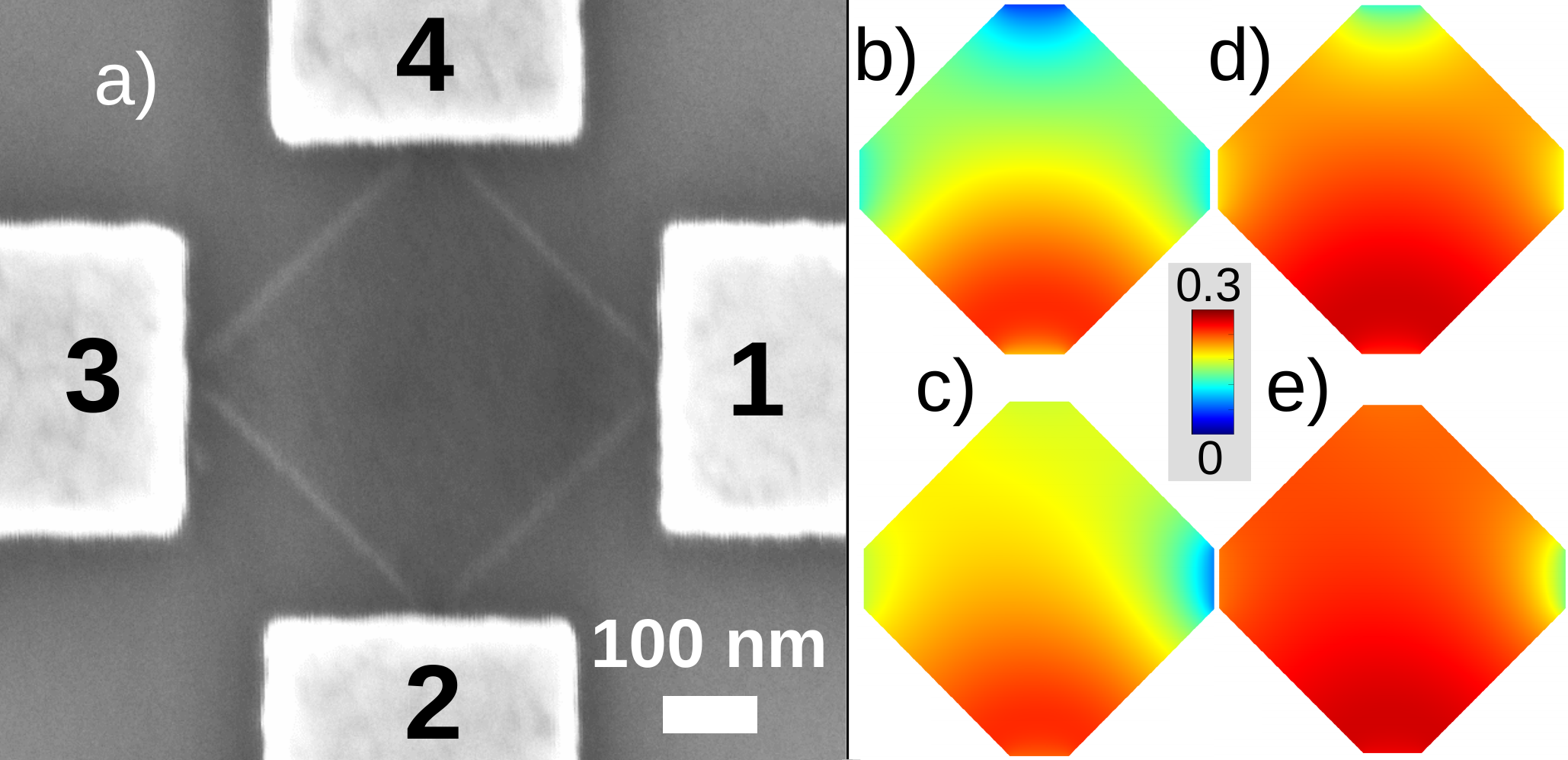}
  \end{center}
\caption{(a) Graphene box sample with Cr/Au contacts at the corners. The scale bar indicates 100 nm. The graphene extends under each contact by approx. $1~\mu$m. Biasing is applied via ports 2 and 4, while cross-correlation $S_{13}$ is measured between terminals 1 and 3. $S_{13}$ is measured in three DC biasing situations: A) $V_2=V$ and  $V_1= V_3=V_4=0$ B)  $V_4=V$ and  $V_1= V_2=V_3=0$, and C) $V_2=V_4=V$ and $V_1=V_3=0$, the results of which are marked by $-S_A$, $-S_B$, and $-S_C$, respectively. The Hanbury Brown and Twiss exchange effect is observed in the difference $\Delta S = S_C - S_A - S_B$. (b, c) Non-equilibrium distribution functions in coherent regime described by $\Pi$  (see Eq.~\ref{noneq}) for our diffusive graphene box that is biased from terminal 2 while the three other terminals are grounded at two gate voltage values: far from the charge neutrality point (CNP) ($V_g = -10$~V) (b) and near it ($V_g = +15$~V) (c). (d,e) Equivalent temperature distribution in hot electron regime (see Eq.~\ref{eq_T_distribution}) in the same bias configuration at at $V_g = -10$~V (d) and $V_g = +15$~V (e). The color scale is relative to the applied bias voltage $V$.
}
\label{fig-sample}
\end{figure}

The Pauli principle also influences the cross-correlations of current fluctuations in a diffusive system. 
The cross-correlation of the fluctuations of the currents entering the conductor through terminals $m$ and $n$, $S_{nm}$, is defined by
\begin{equation}
S_{nm}=\int\limits_{-\infty}^{\infty}dt\langle\delta\tilde {I}_n(t)\delta
\tilde {I}_m(0)\rangle
\label{crosscorr}
\end{equation}
where we assume low frequency limit $eV \gg \hbar \omega$ relevant to our experiments. Our sample, shown in Fig.~\ref{fig-sample}\,(a), 
has four terminals, which are the metallic leads attached to the corners of the box.

One can derive a very general expression \cite{Blanter2000} for the cross correlation (Eq.~\ref{crosscorr}) in terms of the scattering matrix of the device $\hat s_{mn}$.
For practical calculations we use an alternative approach based on the solution of Boltzmann-Langevin equation for diffusive electrons inside the box \cite{Blanter1997,Sukhorukov1999}, 
which we outline in the next subsection. The two approaches are equivalent because the scattering matrix
can be expressed in terms of the electronic retarded ($G^R$) and advanced ($G^A$) Green's functions and transmission probabilities of the conducting channels 
of the terminals \cite{Fisher1981}. Performing disorder averaging of the products $G^R G^A$ in the diffusive conductor with the aid of the
standard rules \cite{Aleiner2002}, one can reduce the evaluation of the cross correlations (Eq.~\ref{crosscorr}) to the
solution of the diffusion equation.

Here our main focus is the HBT exchange effect which is probed by measuring the cross correlation of the currents in terminals 1 and 3, denoted by $S_{13}$. The correlations are measured in three DC biasing configurations, namely \textit{A, B} and \textit{C}. In \textit{A} (\textit{B}) configuration terminal 2 (4) is biased with voltage \textit{V} while the other three terminals are grounded. In \textit{C} configuration both terminals 2 and 4 are biased, while 1 and 3 are grounded. The measured current-current cross correlations $S_{13}$ are negative, but we follow the notation $S = -S_{13}$ used in Ref.~\cite{Blanter2000} which has positive sign.  
Finally, we consider the difference
\begin{eqnarray}
\Delta S = S_{C} - S_{A} - S_{B}.
\label{DS13}
\end{eqnarray}  
By obtaining the exchange correction factor $\Delta S$ from the measured electronic shot noise, our measurement essentially repeats the original HBT experiment performed with photons \cite{R.HanburyBrown1956}. 
For distinguishable non-interacting particles the noises coming from different sources are additive and the combination (Eq.~\ref{DS13}) equals to zero ($\Delta S=0$).
However, since the electrons are indistinguishable and obey Fermi statistics one finds that $\Delta S\not=0$. 
In theory the cross correlation $S_{13}$ is given by the sum of partial contributions $S_{13}^{\alpha\beta}$
containing the combinations of the distribution functions in the leads of the form $f_\alpha(1-f_\beta)+(1-f_\alpha)f_\beta$,
$S_{13}=\sum_{\alpha,\beta=1}^4 S_{13}^{\alpha\beta}$.
The non-zero HBT exchange correction in Eq.~(\ref{DS13}) originates from the contribution $S_{13}^{24}$ having the form \cite{Blanter1997}
\begin{eqnarray}
S_{13}^{24} &=&  \frac{2e^2}{\pi\hbar} \int dE\, \,{\rm Re} \left({\rm Tr}[ \hat s_{41}^\dagger \hat s_{12} \hat s_{23}^\dagger \hat s_{34} ]\right) 
\nonumber\\ && \times\,
[(1-f_2)f_4 +  f_2(1-f_4)].
\end{eqnarray}
Bias configurations \textit{A} and \textit{B} produce the same value for $S_{13}^{24}$, while it vanishes in the bias configuration \textit{C} and at zero temperature 
due to the Pauli principle. Indeed, in this case one finds $(1-f_2)f_4 +  f_2(1-f_4)=0$ since $f_2=f_4=\theta(eV-E)$,
where $\theta(x)$ is the Heaviside step function. Thus, at zero temperature one obtains \cite{Blanter1997}
\begin{eqnarray}
\Delta S &=& 2S_{A,13}^{24} 
\nonumber\\
&=& \frac{4e^2}{\pi\hbar} 
\left\langle \,{\rm Re} \left({\rm Tr}[ \hat s_{41}^\dagger \hat s_{12} \hat s_{23}^\dagger \hat s_{34} ]\right)\right\rangle eV.
\label{DS}
\end{eqnarray}
Here the angular brackets denote averaging over disorder in the diffusive conductor.
The HBT exchange correction (Eqs.~\ref{DS13} and \ref{DS}) can be either positive or negative depending on the system parameters.

As we have mentioned earlier, after disorder averaging quantum interference effects vanish from the HBT exchange noise (Eq.~\ref{DS}).
However, from mathematical point of view one can still consider it as a classical interference 
effect for the distribution function of electrons. Indeed, the distribution function inside the graphene box $f_0$
is the linear combination of the distribution functions in the terminals, see Eq.~(\ref{linear_combination}),
while the noise cross correlation is the quadratic function of it. 
It is well known that the original HBT experiment \cite{R.HanburyBrown1956} can also be interpreted in terms of the interference
of classical waves. The interpretation of our experiment as analogy to optical interference is discussed further in Sec.~VI.

\subsection{Semiclassical analysis}

The non-equilibrium electron transport can be described by Boltzmann-Langevin approach, \cite{Sukhorukov1998,Sukhorukov1999} that provides a simple and transparent interpretation of the theory. In this section
we provide a brief summary of this approach and derive explicit expressions for the noise cross-correlations in terms of measurable parameters.
We account for the effect of finite contact resistances and consider the two regimes -- the regime of the elastic transport and the hot electron regime,
in which electron-electron interaction leads to thermalization of the electrons and a local electronic temperature can be defined. 

Considering the elastic transport regime, in which the electron-electron Coulomb interaction can be ignored, one obtains 
the solution of the Boltzmann equation for the electron distribution function in the form
\begin{eqnarray}
f_0(\varepsilon,{\bm r})=\sum_n\phi_n({\bm r}) f_T(\varepsilon -eV_n),
\label{linear_combination}
\end{eqnarray}
where $\phi_n({\bm r})$ denotes the potential distribution in a diffusive multiterminal conductor corresponding to the bias condition $V_m=\delta_{mn}$.

The noise correlations can be expressed in terms of a function $\Pi$ which describes 
the non-equilibrium state of the biased multiterminal conductor: 
\begin{equation}
\Pi({\bm r}) =2\int d\varepsilon\,f_0(\varepsilon -eV_k,{\bm r})[1-f_0(\varepsilon -eV_l,{\bm r})].
\label{noneq}
\end{equation}
If only one terminal is biased, the function simplifies to $\Pi =e\phi_k (1-\phi_k)|V|$ in the limit $T \rightarrow 0$. With two bias voltages, for example at terminals 2 and 4, one obtains $\Pi =e(\phi_2+\phi_4) (1-(\phi_2+\phi_4) )|V|$. Note that the non-linear dependence of the distribution function ($f_0$) is carried over to dependence on the characteristic function ($\Pi$).
Figs.~\ref{fig-sample}\,(b,c) display the numerically calculated $\Pi$-functions for a graphene box where the contacts are placed in the corners of the box and their effective width is taken as 20\% of the side length $L$. 
The shape of the $\Pi$ function characterizes the diffusion of electrons governed by quantum statistics of fermions.

The noise currents in each terminal can be obtained by integrating the $\Pi$ function. For example, the expression for the noise cross-correlations in a graphene box with perfect contacts  reads
\begin{eqnarray}
S_{ij}= \frac{1}{R_\Box}\int d^2{\bm r}\, \Pi({\bm r})\nabla\phi_i({\bm r})\nabla\phi_j({\bm r}),
\label{Sij}
\end{eqnarray}  
where $R_\Box$ is the sheet resistance of graphene.
In our experimental configuration with finite contact resistances $\phi_n({\bm r})$ exhibit jumps across the contacts, 
which reflect finite voltage drops on them. 
The effect of the contacts on the noise cross-correlations is discussed below.

One can use the elastic approximation for the electron transport if the escape time of an electron out of the graphene quantum dot, $\tau_{\rm esc}$,
is much shorter than the electron-electron energy relaxation time $\tau_{\rm ee}$, i.e. if $\tau_{\rm esc}\ll \tau_{\rm ee}$.
In the opposite case, $\tau_{\rm esc}\gg \tau_{\rm ee}$, hot electron regime becomes relevant.
The time $\tau_{\rm esc}$ is given by the expression
\begin{eqnarray}
\frac{1}{\tau_{\rm esc}}=\frac{\delta_{\rm d}}{4\pi\hbar} \left(\frac{R_q}{R_\Box}+\sum_{k=1}^4 \frac{R_q}{R_k}\right),
\label{tau_esc}
\end{eqnarray}
where $R_q=h/e^2$ is the resistance quantum, $R_k$ are the contact resistances and $\delta_{\rm d}$ is the level spacing in the square graphene dot,
\begin{eqnarray}
\delta_{\rm d}=\frac{\pi\hbar v_0}{L^2 k_F}.
\end{eqnarray}
Here $v_0\approx 10^6$~m/s is the speed of electrons in graphene and $k_F$ is the Fermi wave vector.
The electron-electron relaxation time is estimated as \cite{voutilainen2010},
\begin{eqnarray}
\frac{1}{\tau_{\rm ee}} = \frac{2R_\Box}{R_q}\frac{k_BT_e}{\hbar}\ln\left[\frac{R_q^3}{64R_\Box^3}\frac{e^4 k_F}{\hbar v_0 k_BT_e}\right],
\label{tau_ee}
\end{eqnarray} 
where $T_e$ is the average effective temperature of electrons inside the graphene box. The temperature $T_e$ equals to the bath temperature at
low bias voltages applied to the contacts and may grow to higher values $T_e\sim eV$ in the hot electron regime. 
For the parameters of our sample listed in Tab.~\ref{tab_cond} we find that the times (Eqs.\ \ref{tau_esc} and \ref{tau_ee}) weakly depend on the gate voltage. The escape time
approximately takes the value $\tau_{\rm esc}\approx 1$~ps, 
while the electron-electron relaxation time (Eq.\ \ref{tau_ee}) may change from $\tau_{\rm ee}\sim 50$~ps at the bath temperature $T_e=20$~mK
to much shorter values $\tau_{\rm ee}\ll \tau_{\rm esc}$ at high bias. Thus we expect our sample to be in an intermediate regime
between ballistic and hot electron transport.

In presence of the inelastic electron-electron scattering the shape of the $\Pi$-function changes. 
The kinetic equation for the distribution function can be relatively easily found in the hot electron regime
$\tau_{\rm ee}\ll \tau_{\rm esc} \ll \tau_{\rm e-ph}$, where $\tau_{\rm e-ph}$ is the electron-phonon relaxation time.
In this regime the electron distribution function has the equilibrium Fermi-Dirac form with coordinate dependent
electron temperature, which differs from the temperature of the substrate.

The function $\Pi$ (Eq.~\ref{noneq}) can be expressed in terms of the characteristic functions $\phi_j({\bm r})$ both in
the elastic and the hot electron regimes. Performing this analysis and generalizing the expression (Eq.~\ref{Sij})
to case of finite contact resistances, we derive explicit expressions for  the cross-correlation of the noises in terms
of the experimentally measurable parameters. 
Assuming that the electron transport is fully elastic and considering low temperature (or high bias) limit $k_{\mathrm{B}}T_0\ll eV$ relevant to our 
experiment, we find
\begin{eqnarray}
&& S_{ij}   =  \sum_{k=1}^4 G_{ik}G_{jk}R_k^2 {\cal S}_k 
\nonumber\\ &&
+\,  \sum_{k,l=1}^4 \frac{e|V_k-V_l|}{R_\Box}
\int d^2{\bm r}\, \phi_k({\bm r})\phi_l({\bm r})\nabla\phi_i({\bm r})\nabla\phi_j({\bm r}).\hspace{6mm}
\label{elastic}
\end{eqnarray}
Here $G_{ik}$ are the elements of the conductance matrix, which describe the combined effect of all contact resistances
and the inner part of the graphene box, $R_k$ are contact resistances,
and ${\cal S}_k$ are the local noise sources of the contacts
evaluated under the assumption of fixed potential of the graphene box. The latter have the form
\begin{eqnarray}
{\cal S}_k &=& -e\sum_{l=1}^4 G_{kl}|V_k-V_l|
\nonumber\\ &&
+\,\frac{e(1-F_k)R_k}{2}\sum_{p,l=1}^4 G_{kl}G_{kp}|V_p-V_l|.
\end{eqnarray}
Here  $F_k$ is the Fano factor of the $k$-th contact.
The integral in the last term of Eq. (\ref{elastic}) runs over the inner part of the graphene box excluding
the corner areas, to which the metallic leads are attached.

In the hot electron regime and for $k_{\mathrm{B}}T_0\ll eV$ the cross-correlation takes the form
\begin{eqnarray}
S_{ij} &=&\sum_{k=1}^4 G_{ik}G_{jk}R_k^2{\cal S}_k 
\nonumber\\ &&
+\, \frac{2}{R_\Box}\int d^2{\bm r}\, T_e({\bm r})\nabla\phi_i({\bm r})\nabla\phi_j({\bm r}).
\label{hot_electron} 
\end{eqnarray}
Here $T_e({\bm r})$ is the coordinate dependent electronic temperature inside the graphene box
given by the expression
\begin{eqnarray}
T_e({\bm r})=\sqrt{\frac{3e^2}{2\pi^2}\sum_{p,l=1}^4 \phi_p({\bm r})\phi_l({\bm r})(V_p-V_l)^2},
\label{eq_T_distribution}
\end{eqnarray}
${\cal S}_k$ are again the local junction noise sources, which now take the form
\begin{eqnarray}
{\cal S}_k &=& \frac{F_kT_k}{R_k}\ln\left[2+2\cosh\left(\frac{eR_k\sum_{l=1}^4 G_{kl}|V_k-V_l|}{T_k}\right)\right]
\nonumber\\ &&
+\, \frac{(1-F_k)T_k}{R_k},
\end{eqnarray}
and $T_k$ are the electronic temperatures inside the box close to the contacts,
\begin{eqnarray}
&& T_k=\frac{\sqrt{3}eR_k}{\sqrt{2}\pi}
\nonumber\\ && \times\,
\sqrt{\sum_{p,l=1}^4 G_{kp}G_{kl}(V_k-V_l)^2 - \sum_{l=1}^4 \frac{2G_{kl}}{R_k}(V_k-V_l)^2}.\hspace{5mm}
\end{eqnarray}

\subsection{Conductance}
\begin{figure}[!ht]
\centering
\includegraphics[width=1\linewidth]{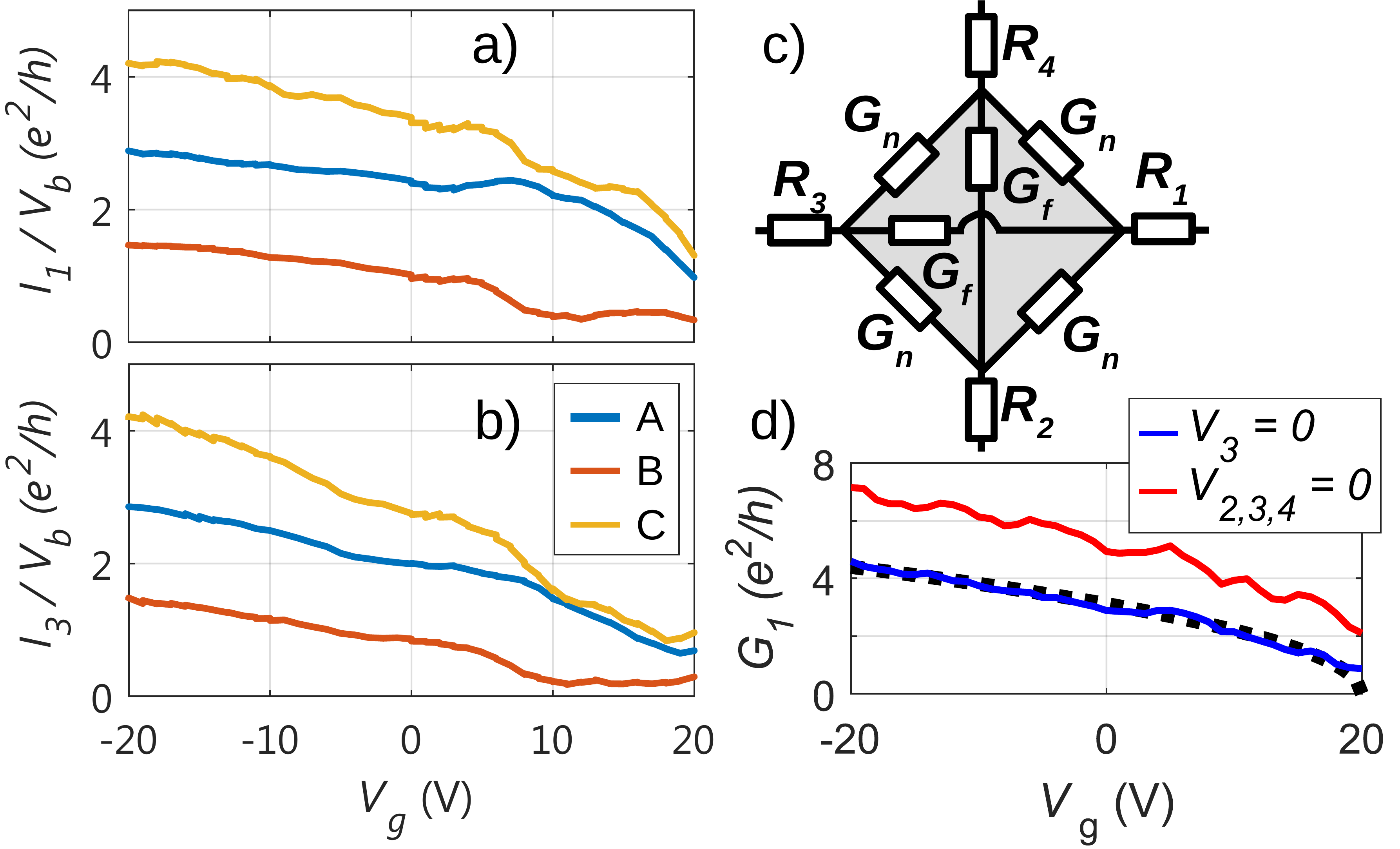}
\caption{(a, b) Ratio of measured current and bias voltage at contacts 1 (a) and 3 (b) in bias configurations A, B and C. (c) Schematic illustrating the division of the system into contacts and uniformly conducting graphene. The geometry used in our diffusive model yields $G_n = 1.6\,G_f$ and $R_\square = 0.165\,G_f^{-1}$. (d) Measured conductance as a function of gate voltage at contact 1 ($G_1=I_1/V_1$) with other contacts grounded (red curve) and with contact 3 grounded and the other two floating ($I_2=I_4=0$, blue curve). The dotted black line shows fit to theoretical conductance of two constrictions (Eq.~\ref{eq-G}) and central region (with $R_1 + R_3 = 0.5R_{\mathrm{tot}}$) in series with $W = 50$~nm, $c_0 = 0.90$ and $k_{\mathrm{F}}$ set to its theoretical value ($V_{g,\mathrm{CNP}}$ set to +20~V).  \label{fig-cond}}
\end{figure}

As described above, the conductances of graphene and contacts are parameters in our numerical noise model. Therefore, we use the measured conductances shown in Fig.~\ref{fig-cond}\,(a,b,d) as a starting point for the numerical noise calculations.

The measured conductances are used to construct a 4-by-4 conductance matrix for the whole system ($\mathbf{G}$) which is then divided to central graphene part ($\mathbf{\tilde{G}}$) with uniform conductivity and contact resistances (diagonal matrix $\mathbf{R}$), satisfying $\mathbf{G} = \mathbf{\tilde{G}}(\mathbf{\tilde{G}} + \mathbf{R}^{-1})^{-1}\mathbf{R}^{-1}$. The division is illustrated in Fig.~\ref{fig-cond}\,(c). Since the magnitude of graphene resistance in this division is largely arbitrary, the graphene resistance value is based on theoretical sheet conductivity at given gate voltage value. The resistances are listed in Table~\ref{tab_cond}.

\begin{table}[ht]
\centering
\begin{tabular}{|c||c|c|c|c|c|}
\hline
$V_g$  & $R_{1}$ & $R_{2}$ & $R_{3}$ & $R_{4}$ & $ R_\square $ \\
\hline
\hline
$-10$~V  & 1.39~k$\Omega$ & 1.50~k$\Omega$ & 1.66~k$\Omega$ & 5.38~k$\Omega$ & 1.65~k$\Omega$  \\
\hline
$+15$~V & 1.59~k$\Omega$ & 4.47~k$\Omega$ & 8.00~k$\Omega$ & 36.4~k$\Omega$ & 3.70~k$\Omega$ \\
\hline
\end{tabular}

\caption{\label{tab_cond}Contact resistances ($R_{i}$) and graphene sheet resistivity ($R_\square$) used in the numerical calculations far from CNP ($V_g = -10$~V) and near it ($V_g = +15$~V).}
\end{table}

It can be seen that the contacts 1-3 have comparable resistances far from the CNP while contact 4 has higher resistance. The differences between the contacts become more significant when approaching the CNP. 

The relatively high contact resistances ($R_i$) are to a large extent explained by narrow regions in the graphene, which can be thought as graphene nanoconstrictions \cite{Terres2016, KrishnaKumar2017, Clerico2018}. Therefore, their effect is briefly studied below. The conductance of such nanoconstriction is given by:

\begin{equation}
    G_{\mathrm{GNC}} = \frac{4 e^2}{h} \frac{c_0 W k_{\mathrm{F}}}{\pi}, \label{eq-G}
\end{equation}
where $c_0 (\leq 1)$ is related to edge roughness ($c_0 < 1$ for rough edges), $W$ is the width of the constriction and $k_{\mathrm{F}} = \sqrt{\pi n}$ is the Fermi wave vector in graphene. For 300~nm gate oxide $n \approx |V_g-V_{g,\mathrm{CNP}}| \times 7.2 \times 10^{10}~\mathrm{cm}^{-2}$, where $V_{g,\mathrm{CNP}}$ is the gate voltage corresponding to the charge neutrality point \cite{DasSarma2011}.

The blue curve in Fig.~\ref{fig-cond}\,(d) shows the measured conductance as a function of gate voltage between terminals 1 and 3 (with 2 and 4 floating, i.e.\ $I_2=I_4=0$) and fit to the constriction model (Eq.~\ref{eq-G}) as $G_1 = (R_1 + R_c + R_3)^{-1} = G_{\mathrm{GNC}}/4$, where the resistance of the central region, $R_c = (G_f+G_n)^{-1}$, contributes by 50\,\% to the total resistance (far from CNP) according to our conductance model. In the calculation we use $W = 50$~nm (estimated from SEM image) and set the value of $k_{\mathrm{F}}$ to its theoretical value. 
Good agreement is found by setting the edge roughness parameter $c_0 \approx 0.90$, which is close to the previously reported experimental values 0.56 \cite{Terres2016} and 0.74 \cite{Clerico2018}. One may also deduce the number of conduction channels in the contacts ($=Wk_{\mathrm{F}}/\pi$), which becomes $\sim$7 far from CNP ($V_g = -10$~V) and $\sim$3 near it ($V_g = +15$~V) in our device.  
However, it should be noted that the presence of increased carrier density due to proximity of metallic contacts can increase $k_{\mathrm{F}}$, leading to smaller $c_0$, and therefore the obtained parameter values are only estimates. Also, here we assume $R_i = G_{\mathrm{GNC}} ^{-1}$, ignoring possible other contributions to contact resistance. We note that the non-zero conductance near the CNP is most probably caused by doping from contacts and impurities.

\subsection{Numerical calculations}\label{sec-numerical}
We base our numerical calculations on the coherent and hot electron models described above. While the contact contributions are readily obtainable from the first terms of Eqs.~(\ref{elastic}) and (\ref{hot_electron}), the graphene terms are calculated numerically. We find the four characteristic functions $\phi_k$ by numerically solving \footnote{Comsol Multiphysics was used for the calculation.} the diffusion equation $\nabla \cdot \hat{\sigma} \nabla \phi_k = 0$ in a 2D geometry representing the graphene box. The chamfered corner terminals (width 20~\% of box edge) have a constant voltage by setting  $V_m = \delta_{mn}V_0$ (for $\phi_n$) for the whole system, and subtracting the voltage drop in each contact, calculated using $\mathbf{G}$ and $\mathbf{R}$. Thus the corner terminals have Dirichlet boundary condition while the box edges have zero perpendicular flux (Neumann boundary). The values of $\phi_k$ (for $k = \{2,4\}$) and $\nabla \phi_k$ (for $k = \{1,3\}$) are evaluated in a $\sim670 \times 670$ grid for numerical integration.

We obtain the noise cross-correlations $S_{13}$ in bias configurations A, B and C (denoted by $S_{A,B,C}$) and equivalently the autocorrelations $S_{11}$ and $S_{33}$, which are denoted by $\Xi_{A,B,C}$. To compare the calculated and measured results more easily we introduce a dimensionless scaled exchange factor $\Delta S_{\mathrm{scaled}} = \Delta S / (S_A + S_B)$ (and similarly $\Delta\Xi_{\mathrm{scaled}}$). Both quantities are calculated far from CNP (at $V_g = -10$~V) and near it (at $V_g = +15$~V, while $V_{g,\mathrm{CNP}} \approx +20$~V). The contact Fano factors turn out to have only little effect on the end result, and hence we set $F_i = 1/3$; equivalent results are obtained even with values approaching the quantum point contact limit ($F_i = 0$). Since the applied bias is relatively large, we can use the assumption  $k_{\mathrm{B}}T_0\ll eV$ in the hot electron regime.

The calculated distributions of the $\Pi$ function in bias configuration A (and equivalent temperature $T$ for hot electron regime) are presented in Fig.~\ref{fig-sample}\,(b--e). 
It can be seen that the distribution concentrates near the biasing terminal in the coherent regime (b,c) while more uniform distributions are observed in the hot electron regime (d,e). The increased asymmetry of contact resistances near the CNP (c,e) is also reflected to the noise distributions.

\begin{figure}
  \begin{center}
    \leavevmode
\includegraphics[width=1\linewidth]{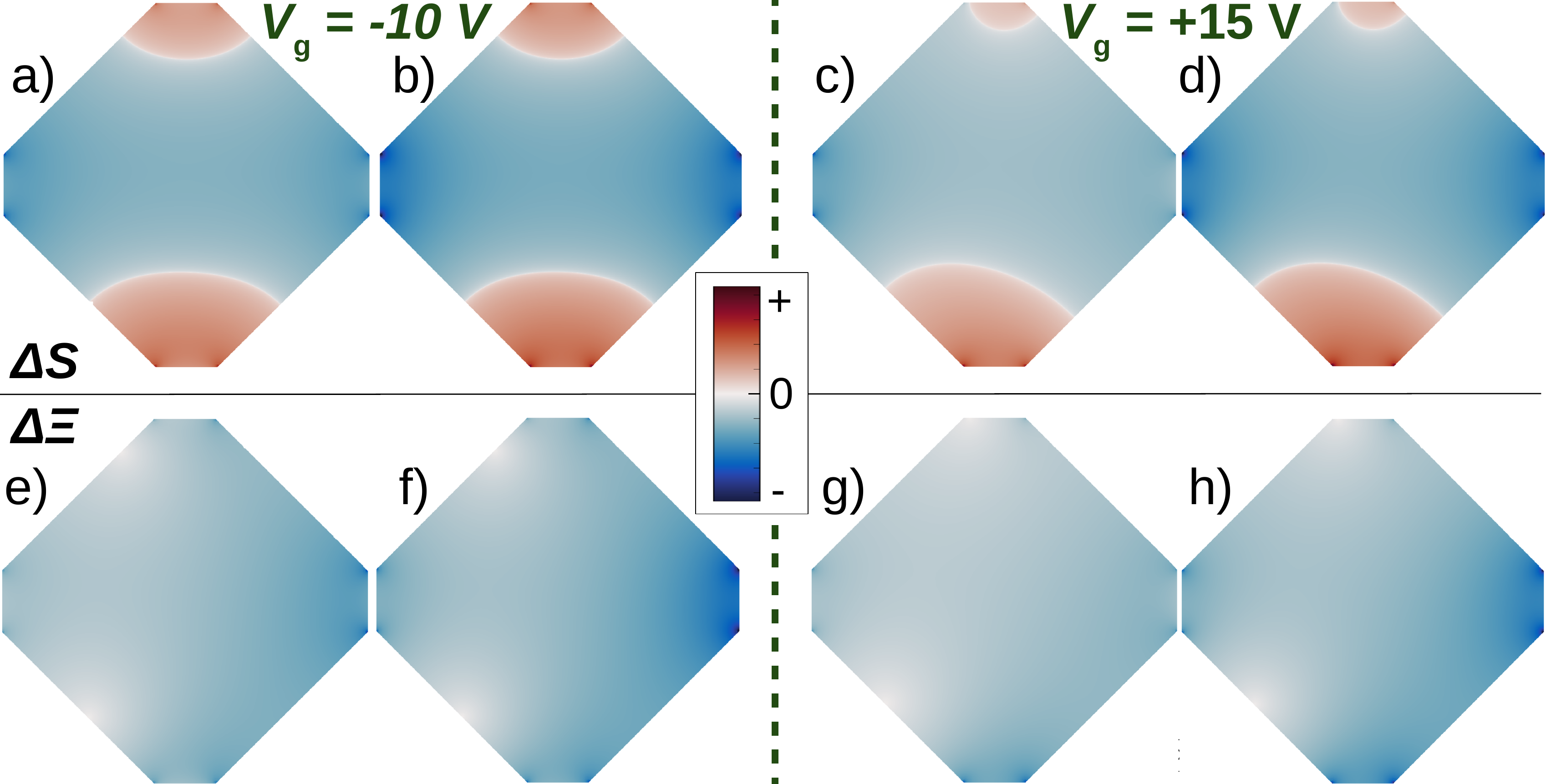}
  \end{center}
\caption{Calculated integrands of $\Delta S$ (top row) and $\Delta \Xi$ probed at terminal 1 on the right (see Fig.~\ref{fig-sample}\,(a) for terminal numbering) (bottom row) at $V_g = -10$~V (left half) and $V_g = +15$~V (right half) with coherent (a,c,e,g) and hot electron models (b,d,f,h). The plots show cube roots of the data to enhance the visual clarity. All values are scaled with their respective integrated $S_A + S_B$ (or $\Xi_A + \Xi_B$) for comparison.}
\label{fig-surfsDS}
\end{figure}

The distributions of the exchange factors $\Delta S$ and $\Delta \Xi$ are shown in Fig.~\ref{fig-surfsDS}. The integrated values of the these quantities are negative, although the distributions of $\Delta S$ have positive contribution near the biasing contacts 2 and 4, where $\nabla\phi_1 \cdot \nabla\phi_3 < 0$. The low conductance of contact 4 reduces the size of the positive region near it, and increased conductance asymmetry at $V_g=+15$~V also clearly increases the asymmetry in the distributions. In general, the difference between the coherent and hot electron regimes appears as small change in the overall level, although the integrated values show larger difference. The values of $\Delta \Xi$ are negative over the whole box, and the distributions are slightly concentrated towards the probing terminal 1 (on the right).

\begin{figure}
  \begin{center}
    \leavevmode
\includegraphics[width=0.99\linewidth]{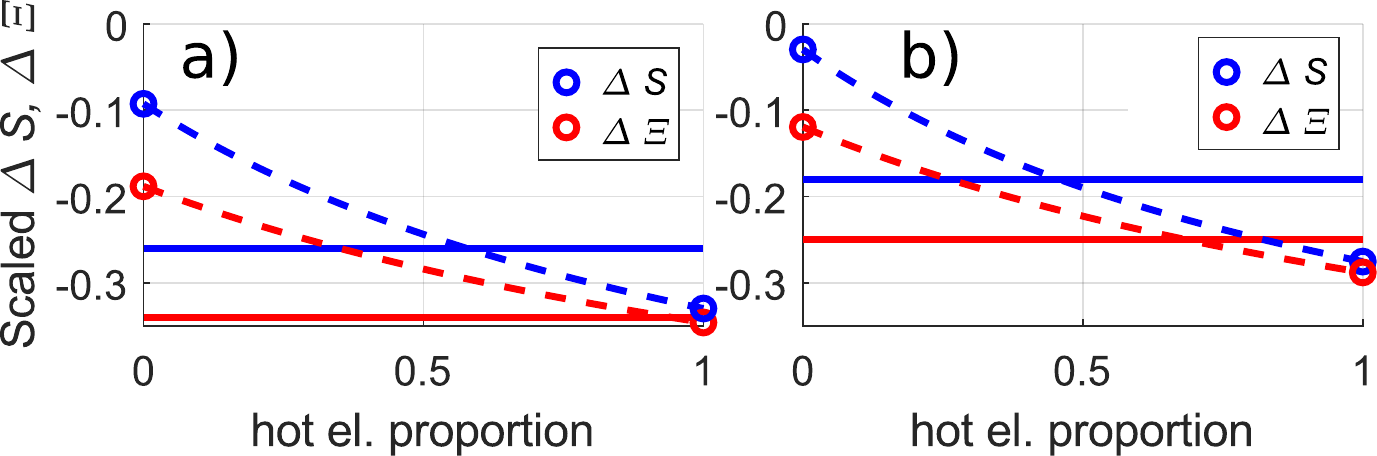}
  \end{center}
\caption{Calculated scaled exchange factors $\Delta S_{\mathrm{scaled}}$ (blue circles) and $\Delta \Xi_{\mathrm{scaled}}$ (red circles) for coherent and hot electron models at (a) $V_g = -10$~V and (b) $V_g = +15$~V. The solid lines indicate approximate experimental results (see Fig.~\ref{fig-deltaS}). The dashed curves connecting the two regimes are calculated from linearly interpolated $S_{A,B,C}$ and $\Xi_{A,B,C}$ between the coherent and hot electron values.}
\label{fig-hot-el-prop}
\end{figure}

The calculated $\Delta S_{\mathrm{scaled}}$ and $\Delta \Xi_{\mathrm{scaled}}$ including the contact and graphene contributions in coherent and hot electron regimes are plotted in Fig.~\ref{fig-hot-el-prop}\,(a) and (b) for far and near the CNP, respectively. The exchange factors are plotted as a function of hot electron proportion: the coherent result is on the left end and hot electron result on the right with a crossover regime between the two extremes. It can be seen that the coherent model results in too weak exchange factors compared to the experiment, while hot electron regime produces too strong $\Delta S_{\mathrm{scaled}}$. Since the experimental results fall between the two regimes, we approximate the coherent - hot electron crossover regime by applying linear interpolation as a function of hot electron contribution to calculated cross- and autocorrelations ($S_{A,B,C}$ and $\Xi_{A,B,C}$) individually and calculate the resulting exchange factors which are shown as dashed lines in Fig.~\ref{fig-hot-el-prop}. A relatively good agreement is obtained at $\sim60$~\% hot electron contribution at $V_g = -10$~V and $\sim50$~\% at $V_g = +15$~V. It should be noted, however, that such interpolation only provides a rough estimate of the behavior in the crossover regime.

\section{Experimental methods}
The sample (see Fig.~\ref{fig-sample}\,(a)) is fabricated from micromechanically cleaved graphene on heavily p-doped substrate with 300~nm gate oxide. The graphene extends under the Cr/Au contact electrodes. The bonding pads are sufficiently small ($150\times150~\mathrm{\mu m^2}$) so that only 10~\% of noise is shunted capacitively to the back-gate electrode.

\begin{figure}
  \begin{center}
    \leavevmode
\includegraphics[width=0.95\linewidth]{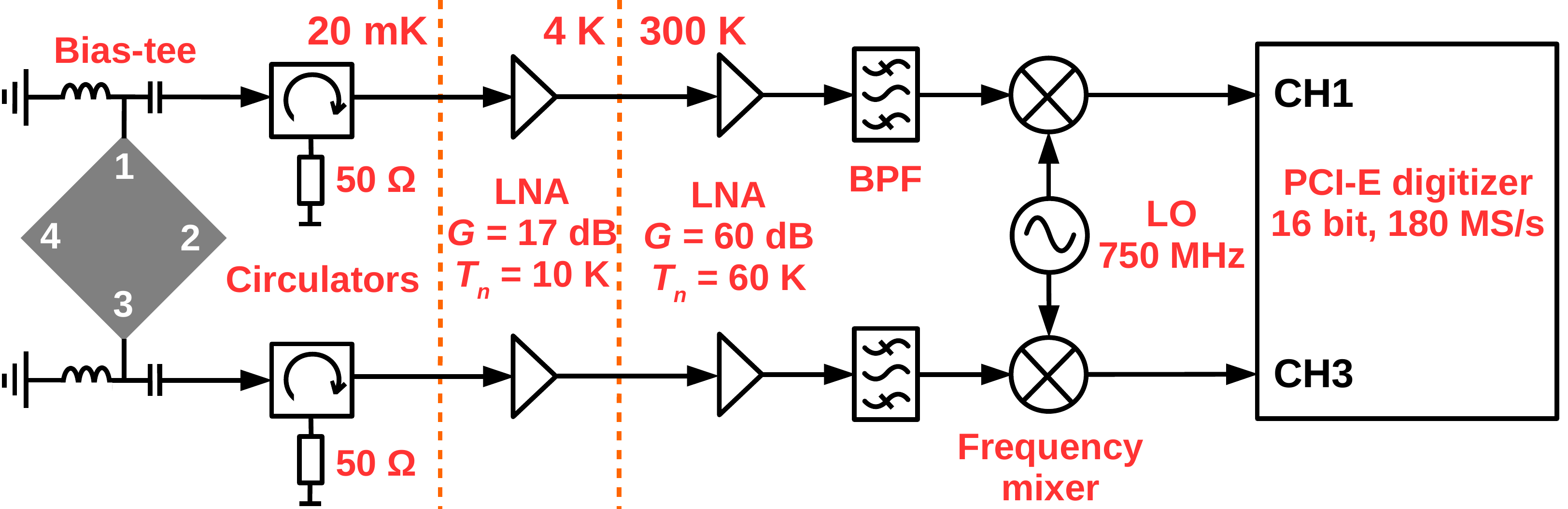}
  \end{center}
\caption{Schematic of the measurement configuration.
}
\label{fig-meas_setup}
\end{figure}

A schematic of the experimental setup is shown in Fig.~\ref{fig-meas_setup}. The experiments are conducted on a BlueFors dry dilution refrigerator at 20~mK. The sample is connected to two high-frequency measurement channels with bias-tees separating the DC and RF paths. Both channels have home-made HEMT low-noise amplifiers (LNA) mounted at 4~K stage \cite{Nieminen2016} reaching system noise temperatures of $\sim$10~K for channel 1 and $\sim$15~K for channel 3. After additional amplification and band-pass filtering (BPF) in room temperature the RF signal is mixed down with a local oscillator (LO) at 750~MHz and digitized at 180 megasamples per second (MS/s) with AlazarTech ATS9642 digitizer connected to PCI-E bus of a desktop computer. The cross- and autocorrelations are calculated from the digitized data using graphics processing unit (GPU) acceleration. Noise power coupling issues were treated along the lines given in Ref. \cite{Danneau2008a}.

\section{Results}
\begin{figure}
  \begin{center}
    \leavevmode
\includegraphics[width=0.95\linewidth]{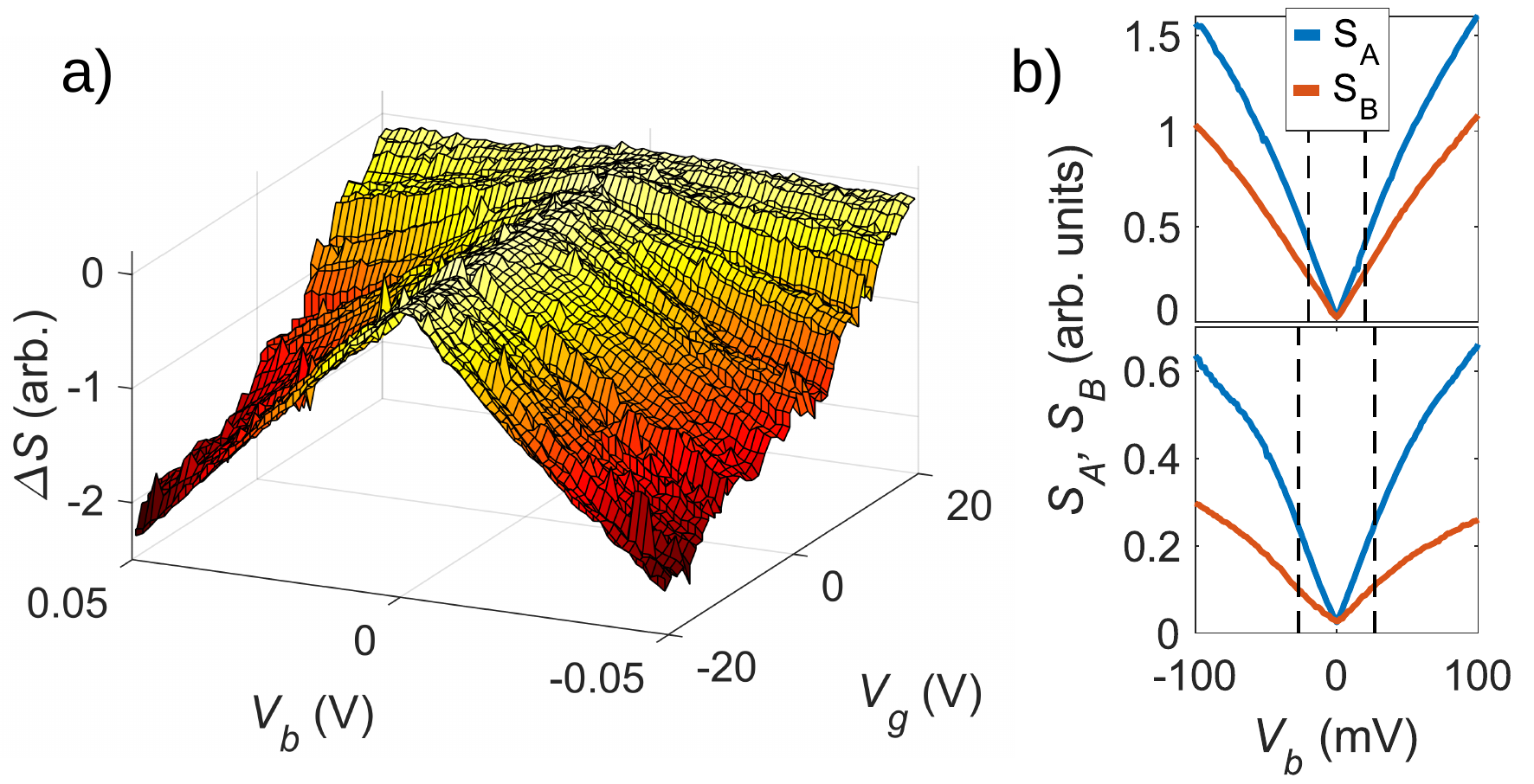}
  \end{center}
\caption{(a) HBT exchange contribution $\Delta S = S_C - S_A - S_B$ measured on the bias vs. gate voltage plane. The linear negative slope at small bias indicates an approximately constant HBT exchange effect $\Delta S /(S_A + S_B)$ as a function of $V_b$. (b) Measured noise cross-correlations $S_A$ and $S_B$ (in arbitrary units) at $V_g = -10$~V (top) and $V_g = +15$~V (bottom). The dashed vertical lines denote the ranges $V_b = -20 ... +20$~mV (for $V_g \leq 0$~V) and $V_b = -27 ... +27$~mV (for $V_g > 0$~V), which are used for linear fits. 
}
\label{fig-deltaS_surf}
\end{figure}

Our cross-correlation results for the HBT exchange term $\Delta S = S_C - S_A - S_B$ are displayed in Fig. \ref{fig-deltaS_surf}\,(a) on the plane spanned by the gate ($V_g$) and bias ($V_b$) voltages; the Dirac point is located around $V_g=+20$ V. At small bias, we observe a clear negative HBT effect; as expected for fermionic diffusion, the $\Delta S$ signal  grows linearly with the bias voltage $V_b$. A suppression of noise due to the interference of mutually incoherent electrons has been observed in an experiment with a ballistic electron beam splitter \cite{Liu1998}. Our results demonstrate that this effect is also observable in mesoscopic diffusive conductors.

The value for scaled $\Delta S$ was calculated by making linear fits to the measured noise cross-correlation $S_{A,B,C}$ vs. bias voltage at $V_b = -20 ... 0$ and $0 ... +20$~mV for $V_g \leq 0$~V. The lower conductance near the CNP increases the variance of the data, and therefore a wider range of $V_b$ between $\pm27$~mV is used for $V_g > 0$~V. The data are linear within those intervals, as shown in Fig.~\ref{fig-deltaS_surf}\,(b), although some deviation emerges when approaching the Dirac point due to lower conductance. The use of smaller intervals in fitting increased the variance of the resulting $\Delta S$ due to statistical errors, but the average values remained the same. Therefore, our fits can be considered as small-bias extrapolation to zero bias. We calculate the exchange factors for negative and positive $V_b$ separately to see the scattering of the data.

\begin{figure}
  \begin{center}
    \leavevmode
\includegraphics[width=0.8\linewidth]{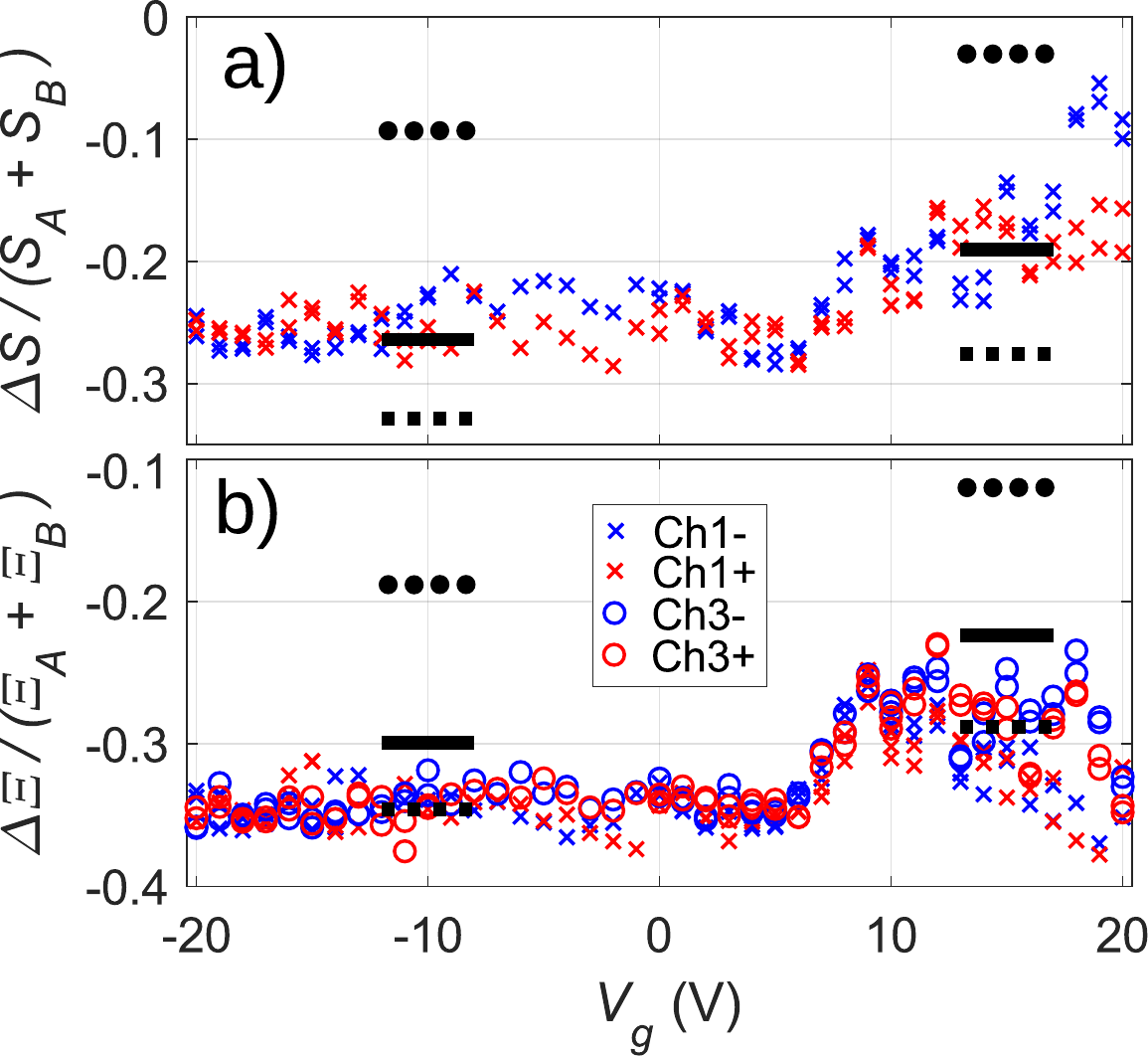}
  \end{center}
\caption{(a) Scaled cross-correlation HBT exchange effect $\Delta S /(S_A + S_B)$ with blue (red) markers corresponding to negative (positive) $V_b$. The black lines dotted with circles and squares denote the results of diffusive calculation in coherent and hot electron regimes, respectively (see Fig.~\ref{fig-hot-el-prop}) while the solid black lines are interpolations in crossover regime between coherent and hot electron regimes.
(b) Scaled autocorrelation HBT exchange effect measured at terminals 1 and 3. 
For the linear interpolations, the hot electron contributions are taken as 60~\% for $V_g = -10$~V and 50~\% for $V_g = +15$~V. 
}
\label{fig-deltaS}
\end{figure}

The determined $\Delta S_{\mathrm{scaled}}$ as a function of gate voltage is shown in Fig.~\ref{fig-deltaS}\,(a) together with numerical results of the coherent and hot electron models ($V_g\approx-10$~V) and close to ($V_g\approx+15$~V) the CNP. The data are scattered mainly due to statistical errors, although a clear trend in $\Delta S_{\mathrm{scaled}}$ can be seen: the effect stays rather constant between $-20~\mathrm{V} \leq V_g  \lesssim +5~\mathrm{V}$ and tends linearly towards zero when approaching the Dirac point. The increased scattering of the data at large $V_g$ is due to smaller absolute values of noise, as seen in Fig.~\ref{fig-deltaS_surf}, and resulting statistical error.

In addition to cross-correlations, we determined the scaled exchange factors $\Delta\Xi_{\mathrm{scaled}}$ for the measured noise in individual channels (autocorrelation). The obtained $\Delta\Xi_{\mathrm{scaled}}$ are shown in Fig.~\ref{fig-deltaS}\,(b) together with the results of the numerical model. $\Delta\Xi_{\mathrm{scaled}}$ shows similar constant behavior below $V_g \approx +5$~V as $\Delta S_{\mathrm{scaled}}$, but the slope approaching the CNP is steeper and the scattering near CNP is significant.

When comparing the experimentally determined exchange factors to the results from diffusive calculations described in Sec.~\ref{sec-numerical}, it can be seen that the experimental values are situated between the coherent and hot electron results (black lines dashed with circles and squares, respectively, in Fig.~\ref{fig-deltaS}). For autocorrelation, the experimental results agree well with the hot electron model, but for cross-correlation the hot electron model results in too strong HBT exhange effect. For best overall agreement, we obtain linearly interpolated values in the crossover regime (see Fig.~\ref{fig-hot-el-prop}) with 60~\% hot electron contribution far from CNP and 50~\% close the CNP. The interpolated values are shown as solid black lines in Fig.~\ref{fig-deltaS}. As already mentioned, however, such linear interpolation gives an inadequate picture of the crossover regime, and therefore the interpolated values should only be considered as rough estimates.


\section{Discussion}

There are several ways to construct a model for a graphene box. One of the simplest is the chaotic quantum dot described by a single distribution function \cite{VanLangen1997}. A straightforward generalization of this model is to adopt the semiclassical model and to describe the graphene using a single distribution function governed by contact resistances with an arbitrary Fano factor. This model is in fact quite close to the model employed in Ref.~\cite{Tan2018}. Such a model, lacking voltage variation over the graphene box, was not able to match all the measured quantities $G_{ij}$, $S_{11}$, $S_{33}$, $S_{13}$ properly. First after inclusion of the characteristic potential distributions, a satisfactory agreement could be achieved.

Closest to the present work is our previous experiment with a graphene cross sample with 50~nm nanoribbon arms \cite{Tan2018}. In that graphene cross the HBT effect was characterized by occupation number noise in the nearly ballistic central region and regular diffusive noise in the ribbon arms. The HBT effect far from CNP in the cross sample ($\Delta S_{\mathrm{scaled}}\approx-0.18$) is of comparable magnitude as in the box ($\approx-0.26$), while the HBT effect near the CNP was strongly enhanced in the cross sample (to $\Delta S_{\mathrm{scaled}}\approx-1.5$) but is suppressed in the graphene box. This is in line with the theoretical findings \cite{Blanter1997} that geometrical details of the sample strongly affect the observed HBT exchange effect.

The shot noise in our sample is generated in the central graphene region, as well as the narrow constrictions at the contacts contributing significantly to sample resistance (see Eq.~\ref{elastic}). The contact resistances affect the characteristic potential distribution in the central region, and thus also their asymmetry has significant effect on the noise. We note that, for such small contacts as we have in our sample, the contact capacitance can be regarded negligible (on the basis of Ref.~\cite{Laitinen2016}), and the reactive impedance part at the noise measurement frequency does not bypass the contact resistance, resulting that the DC conductance values are sufficient for the noise circuit analysis.

The fact that the Dirac point in our sample is not well defined (see Figs.~\ref{fig-cond} and \ref{fig-deltaS}) indicates presence of nonuniform doping, possibly due to contributions from fabrication residues, proximity of the contacts, and localized states at the edges. These locally varying doping effects would lead to nonuniform conductance in the regime of charge puddles near the Dirac point. Nonuniform conductance can easily be implemented in our numerical calculations, but this approach was not found exceedingly beneficial, and was given up due to further increase in the already high number of our fitting parameters. In addition, the characteristic statistics of the charge puddles are unknown, making it impossible to justify any specific configuration of non-uniformity in our model.

According to Fig.~\ref{fig-deltaS}, our results are intermediate between coherent and hot electron transport. Theoretically, however, the strength of electron-electron interactions grows as $V^2$ due to the increase in the available scattering states with bias voltage. Therefore, we would expect the electron propagation in our graphene box transform gradually with bias even closer to hot electron regime with decreasing electron-electron scattering length $\ell_{el-el} < L$ \cite{Nagaev1995}. 
 Instead of an increase in the shot noise due to hot electron effects, we find a 5\% decrease in total $F$ ($F=S_I/2eI$) at $V_b = 50$~mV compared with the value deduced using the low bias Fano factor. This decrease is assigned to inelastic scattering, i.e. to the onset of scattering by polar surface modes in graphene/SiO$_2$ system \cite{Chen2008,PolarModes}. Our experimental results do not show any noticeable change in $\Delta S_{\mathrm{scaled}}$ up to bias voltage $V_b = 50$~mV, and hence pure hot electron regime was not fully achieved in our experiments. The independence of our results on bias at $|V_b| \lesssim 50$~mV, suggests that $\Delta S$ and $\Delta \Xi$ might include features inherent to diffusive graphene.

It is instructive to consider the analogy between our experiment and interference experiments in optics. We note that in Eq.~(\ref{Sij}) the function $\Pi$ is multiplied by gradients $\nabla\phi_m$ and $\nabla\phi_n$. 
In analogy with optics, these gradients can be interpreted as distributed detector functions "filtering" the $\Pi$-function. 
They vary smoothly inside the graphene box, which implies that the whole box acts as an "interferometer screen".
In this interpretation the noise cross-correlation is given by an area integral weighted with the geometric response functions. For a simple description, we employ an analogy with a double slit experiment where the incoming intensity $I$ on the detector is determined in three different configurations:  the experiment is performed by closing first  one slit ($I_A$), then the other slit ($I_B$), and finally by keeping both slits open ($I_C$); here the applied electric potential is the analogue of light in the double slit experiment.  In our case, the "detection screen" is the whole graphene box where interference due to $f(1-f)$ takes place at every point. The recorded interference value is an integral provided by the cross correlation measurements (see Eq.~\ref{Sij}) where we take the equivalent of $\Delta I = I_C - I_A - I_B$ (the difference between the actual interference pattern and the two backgrounds), namely $\Delta S = S_C - S_A - S_B$. Although, this analogue is illuminating for understanding the setting of our experiment, the underlying effects are two-particle interferences. The correlation effects arise via the competition of the available states in the reservoirs for the outgoing electrons. The non-equilibrium $\Pi$-function (see Fig.~\ref{fig-sample}\,(b,c)) carries this information over to the whole sample. However, the actual phase dependence of the two particle scattering events is averaged out in our diffusive conductor \cite{Blanter1997}.  Due to the lack of phase dependence, we prefer to call our observed results as HBT exchange effects, even though interference by two diffusive wave fronts describes the phenomena in the sense of our analogy.

\section{Conclusions}
We have studied exchange cross correlations in a disordered graphene box. Our experimental results display distinct Hanbury Brown and Twiss  (HBT) exchange correlations, which deviate from the standard predictions of scattering matrix theory. Our results indicate that the finite contact resistances significantly affect the noise cross-correlations in a diffusive system. The values of experimentally determined HBT exchange effects fall between calculated values for coherent and hot electron models, indicating either the presence of bias-independent crossover regime or intrinsic behavior of diffusive graphene which is not captured by standard model for diffusive systems. The overall picture is the same for both near and far from the Dirac point, although the low carrier density near the CNP leads to further deviations between the model and experimental data. 

\section*{Acknowledgements}
We thank Florian Libisch and Stefan Rotter for fruitful discussions. This work was supported by the Academy of Finland projects 314448 (BOLOSE), 310086 (LTnoise) and 312295 (CoE, Quantum Technology Finland) as well as by ERC (grant no. 670743). This research project utilized the Aalto University OtaNano/LTL infrastructure which is part of European Microkelvin Platform. T.E.\ is grateful to Vilho, Yrj{\"o} and Kalle V{\"a}is{\"a}l{\"a} Foundation for scholarship. The work of G.B.L.\ was supported by Aalto University School of Science Visiting Professor grant, the Government of the Russian Federation (Agreement No. 05.Y09.21.0018), by the RFBR Grant No. 17-02-00396A, and Foundation for the Advancement
of Theoretical Physics and Mathematics "BASIS".

\bibliography{lahteet3}

\begin{thebibliography}{46}%
\makeatletter
\providecommand \@ifxundefined [1]{%
 \@ifx{#1\undefined}
}%
\providecommand \@ifnum [1]{%
 \ifnum #1\expandafter \@firstoftwo
 \else \expandafter \@secondoftwo
 \fi
}%
\providecommand \@ifx [1]{%
 \ifx #1\expandafter \@firstoftwo
 \else \expandafter \@secondoftwo
 \fi
}%
\providecommand \natexlab [1]{#1}%
\providecommand \enquote  [1]{``#1''}%
\providecommand \bibnamefont  [1]{#1}%
\providecommand \bibfnamefont [1]{#1}%
\providecommand \citenamefont [1]{#1}%
\providecommand \href@noop [0]{\@secondoftwo}%
\providecommand \href [0]{\begingroup \@sanitize@url \@href}%
\providecommand \@href[1]{\@@startlink{#1}\@@href}%
\providecommand \@@href[1]{\endgroup#1\@@endlink}%
\providecommand \@sanitize@url [0]{\catcode `\\12\catcode `\$12\catcode
  `\&12\catcode `\#12\catcode `\^12\catcode `\_12\catcode `\%12\relax}%
\providecommand \@@startlink[1]{}%
\providecommand \@@endlink[0]{}%
\providecommand \url  [0]{\begingroup\@sanitize@url \@url }%
\providecommand \@url [1]{\endgroup\@href {#1}{\urlprefix }}%
\providecommand \urlprefix  [0]{URL }%
\providecommand \Eprint [0]{\href }%
\providecommand \doibase [0]{https://doi.org/}%
\providecommand \selectlanguage [0]{\@gobble}%
\providecommand \bibinfo  [0]{\@secondoftwo}%
\providecommand \bibfield  [0]{\@secondoftwo}%
\providecommand \translation [1]{[#1]}%
\providecommand \BibitemOpen [0]{}%
\providecommand \bibitemStop [0]{}%
\providecommand \bibitemNoStop [0]{.\EOS\space}%
\providecommand \EOS [0]{\spacefactor3000\relax}%
\providecommand \BibitemShut  [1]{\csname bibitem#1\endcsname}%
\let\auto@bib@innerbib\@empty
\bibitem [{\citenamefont {Kogan}(1996)}]{Kogan1996}%
  \BibitemOpen
  \bibfield  {author} {\bibinfo {author} {\bibfnamefont {S.}~\bibnamefont
  {Kogan}},\ }\href {https://doi.org/10.1017/CBO9780511551666} {\emph {\bibinfo
  {title} {{Electronic Noise and Fluctuations in Solids}}}}\ (\bibinfo
  {publisher} {Cambridge University Press},\ \bibinfo {address} {Cambridge},\
  \bibinfo {year} {1996})\BibitemShut {NoStop}%
\bibitem [{\citenamefont {Blanter}\ and\ \citenamefont
  {B{\"{u}}ttiker}(2000)}]{Blanter2000}%
  \BibitemOpen
  \bibfield  {author} {\bibinfo {author} {\bibfnamefont {Y.~M.}\ \bibnamefont
  {Blanter}}\ and\ \bibinfo {author} {\bibfnamefont {M.}~\bibnamefont
  {B{\"{u}}ttiker}},\ }\bibfield  {title} {\bibinfo {title} {{Shot noise in
  mesoscopic conductors}},\ }\href
  {https://doi.org/10.1016/S0370-1573(99)00123-4} {\bibfield  {journal}
  {\bibinfo  {journal} {Phys. Rep.}\ }\textbf {\bibinfo {volume} {336}},\
  \bibinfo {pages} {1} (\bibinfo {year} {2000})}\BibitemShut {NoStop}%
\bibitem [{\citenamefont {Martin}(2005)}]{Martin2005}%
  \BibitemOpen
  \bibfield  {author} {\bibinfo {author} {\bibfnamefont {T.}~\bibnamefont
  {Martin}},\ }\bibfield  {title} {\bibinfo {title} {{Noise in mesoscopic
  physics}},\ }in\ \href {https://doi.org/10.1016/S0924-8099(05)80047-2} {\emph
  {\bibinfo {booktitle} {Les Houches Summer Sch. Proc.}}},\ Vol.~\bibinfo
  {volume} {81},\ \bibinfo {editor} {edited by\ \bibinfo {editor}
  {\bibfnamefont {H.}~\bibnamefont {Bouchiat}}, \bibinfo {editor}
  {\bibfnamefont {Y.}~\bibnamefont {Gefen}}, \bibinfo {editor} {\bibfnamefont
  {S.}~\bibnamefont {Gu{\'{e}}ron}}, \bibinfo {editor} {\bibfnamefont
  {G.}~\bibnamefont {Montambaux}},\ and\ \bibinfo {editor} {\bibfnamefont
  {J.}~\bibnamefont {Dalibard}}}\ (\bibinfo  {publisher} {Elsevier Ltd},\
  \bibinfo {year} {2005})\ Chap.~\bibinfo {chapter} {5}, pp.\ \bibinfo {pages}
  {283--359}\BibitemShut {NoStop}%
\bibitem [{\citenamefont {Lesovik}\ and\ \citenamefont
  {Sadovskyy}(2011)}]{Lesovik2011}%
  \BibitemOpen
  \bibfield  {author} {\bibinfo {author} {\bibfnamefont {G.~B.}\ \bibnamefont
  {Lesovik}}\ and\ \bibinfo {author} {\bibfnamefont {I.~A.}\ \bibnamefont
  {Sadovskyy}},\ }\bibfield  {title} {\bibinfo {title} {{Scattering matrix
  approach to the description of quantum electron transport}},\ }\href
  {https://doi.org/10.3367/UFNe.0181.201110b.1041} {\bibfield  {journal}
  {\bibinfo  {journal} {Physics-Uspekhi}\ }\textbf {\bibinfo {volume} {54}},\
  \bibinfo {pages} {1007} (\bibinfo {year} {2011})}\BibitemShut {NoStop}%
\bibitem [{\citenamefont {Liu}\ \emph {et~al.}(1998)\citenamefont {Liu},
  \citenamefont {Odom}, \citenamefont {Yamamoto},\ and\ \citenamefont
  {Tarucha}}]{Liu1998}%
  \BibitemOpen
  \bibfield  {author} {\bibinfo {author} {\bibfnamefont {R.~C.}\ \bibnamefont
  {Liu}}, \bibinfo {author} {\bibfnamefont {B.}~\bibnamefont {Odom}}, \bibinfo
  {author} {\bibfnamefont {Y.}~\bibnamefont {Yamamoto}},\ and\ \bibinfo
  {author} {\bibfnamefont {S.}~\bibnamefont {Tarucha}},\ }\bibfield  {title}
  {\bibinfo {title} {{Quantum interference in electron collision}},\ }\href
  {https://doi.org/10.1038/34611} {\bibfield  {journal} {\bibinfo  {journal}
  {Nature}\ }\textbf {\bibinfo {volume} {391}},\ \bibinfo {pages} {263}
  (\bibinfo {year} {1998})}\BibitemShut {NoStop}%
\bibitem [{\citenamefont {Oliver}(1999)}]{Oliver1999}%
  \BibitemOpen
  \bibfield  {author} {\bibinfo {author} {\bibfnamefont {W.~D.}\ \bibnamefont
  {Oliver}},\ }\bibfield  {title} {\bibinfo {title} {{Hanbury Brown and
  Twiss-Type Experiment with Electrons}},\ }\href
  {https://doi.org/10.1126/science.284.5412.299} {\bibfield  {journal}
  {\bibinfo  {journal} {Science}\ }\textbf {\bibinfo {volume} {284}},\ \bibinfo
  {pages} {299} (\bibinfo {year} {1999})}\BibitemShut {NoStop}%
\bibitem [{\citenamefont {Henny}\ \emph {et~al.}(1999)\citenamefont {Henny},
  \citenamefont {Oberholzer}, \citenamefont {Strunk}, \citenamefont {Heinzel},
  \citenamefont {Ensslin}, \citenamefont {Holland},\ and\ \citenamefont
  {Sch{\"{o}}nenberger}}]{Henny1999}%
  \BibitemOpen
  \bibfield  {author} {\bibinfo {author} {\bibfnamefont {M.}~\bibnamefont
  {Henny}}, \bibinfo {author} {\bibfnamefont {S.}~\bibnamefont {Oberholzer}},
  \bibinfo {author} {\bibfnamefont {C.}~\bibnamefont {Strunk}}, \bibinfo
  {author} {\bibfnamefont {T.}~\bibnamefont {Heinzel}}, \bibinfo {author}
  {\bibfnamefont {K.}~\bibnamefont {Ensslin}}, \bibinfo {author} {\bibfnamefont
  {M.}~\bibnamefont {Holland}},\ and\ \bibinfo {author} {\bibfnamefont
  {C.}~\bibnamefont {Sch{\"{o}}nenberger}},\ }\bibfield  {title} {\bibinfo
  {title} {{The Fermionic Hanbury Brown and Twiss Experiment}},\ }\href
  {https://doi.org/10.1126/science.284.5412.296} {\bibfield  {journal}
  {\bibinfo  {journal} {Science}\ }\textbf {\bibinfo {volume} {284}},\ \bibinfo
  {pages} {296} (\bibinfo {year} {1999})}\BibitemShut {NoStop}%
\bibitem [{\citenamefont {Glattli}(2005)}]{Glattli2005}%
  \BibitemOpen
  \bibfield  {author} {\bibinfo {author} {\bibfnamefont {D.~C.}\ \bibnamefont
  {Glattli}},\ }\bibinfo {title} {{Tunneling Experiments in the Fractional
  Quantum Hall Effect Regime}},\ in\ \href
  {https://doi.org/10.1007/3-7643-7393-8_5} {\emph {\bibinfo {booktitle}
  {Quantum Hall Eff. Poincar{\'{e}} Semin. 2004}}},\ \bibinfo {editor} {edited
  by\ \bibinfo {editor} {\bibfnamefont {B.}~\bibnamefont {Dou{\c{c}}ot}},
  \bibinfo {editor} {\bibfnamefont {V.}~\bibnamefont {Pasquier}}, \bibinfo
  {editor} {\bibfnamefont {B.}~\bibnamefont {Duplantier}},\ and\ \bibinfo
  {editor} {\bibfnamefont {V.}~\bibnamefont {Rivasseau}}}\ (\bibinfo
  {publisher} {Birkh{\"{a}}user Basel},\ \bibinfo {address} {Basel},\ \bibinfo
  {year} {2005})\ pp.\ \bibinfo {pages} {163--197}\BibitemShut {NoStop}%
\bibitem [{\citenamefont {{R. Hanbury Brown}}\ and\ \citenamefont
  {Twiss}(1956)}]{R.HanburyBrown1956}%
  \BibitemOpen
  \bibfield  {author} {\bibinfo {author} {\bibnamefont {{R. Hanbury Brown}}}\
  and\ \bibinfo {author} {\bibfnamefont {R.~Q.}\ \bibnamefont {Twiss}},\
  }\bibfield  {title} {\bibinfo {title} {{Correlation between photons in two
  coherent beams of light}},\ }\href@noop {} {\bibfield  {journal} {\bibinfo
  {journal} {Nature}\ }\textbf {\bibinfo {volume} {177}},\ \bibinfo {pages}
  {27} (\bibinfo {year} {1956})}\BibitemShut {NoStop}%
\bibitem [{\citenamefont {Neder}\ \emph {et~al.}(2007)\citenamefont {Neder},
  \citenamefont {Ofek}, \citenamefont {Chung}, \citenamefont {Heiblum},
  \citenamefont {Mahalu},\ and\ \citenamefont {Umansky}}]{Neder2007}%
  \BibitemOpen
  \bibfield  {author} {\bibinfo {author} {\bibfnamefont {I.}~\bibnamefont
  {Neder}}, \bibinfo {author} {\bibfnamefont {N.}~\bibnamefont {Ofek}},
  \bibinfo {author} {\bibfnamefont {Y.}~\bibnamefont {Chung}}, \bibinfo
  {author} {\bibfnamefont {M.}~\bibnamefont {Heiblum}}, \bibinfo {author}
  {\bibfnamefont {D.}~\bibnamefont {Mahalu}},\ and\ \bibinfo {author}
  {\bibfnamefont {V.}~\bibnamefont {Umansky}},\ }\bibfield  {title} {\bibinfo
  {title} {{Interference between two indistinguishable electrons from
  independent sources}},\ }\href {https://doi.org/10.1038/nature05955}
  {\bibfield  {journal} {\bibinfo  {journal} {Nature}\ }\textbf {\bibinfo
  {volume} {448}},\ \bibinfo {pages} {333} (\bibinfo {year}
  {2007})}\BibitemShut {NoStop}%
\bibitem [{\citenamefont {Blanter}\ and\ \citenamefont
  {B{\"{u}}ttiker}(1997)}]{Blanter1997}%
  \BibitemOpen
  \bibfield  {author} {\bibinfo {author} {\bibfnamefont {Y.~M.}\ \bibnamefont
  {Blanter}}\ and\ \bibinfo {author} {\bibfnamefont {M.}~\bibnamefont
  {B{\"{u}}ttiker}},\ }\bibfield  {title} {\bibinfo {title} {{Shot-noise
  current-current correlations in multiterminal diffusive conductors}},\ }\href
  {https://doi.org/10.1103/PhysRevB.56.2127} {\bibfield  {journal} {\bibinfo
  {journal} {Phys. Rev. B}\ }\textbf {\bibinfo {volume} {56}},\ \bibinfo
  {pages} {2127} (\bibinfo {year} {1997})}\BibitemShut {NoStop}%
\bibitem [{\citenamefont {Sukhorukov}\ and\ \citenamefont
  {Loss}(1999)}]{Sukhorukov1999}%
  \BibitemOpen
  \bibfield  {author} {\bibinfo {author} {\bibfnamefont {E.~V.}\ \bibnamefont
  {Sukhorukov}}\ and\ \bibinfo {author} {\bibfnamefont {D.}~\bibnamefont
  {Loss}},\ }\bibfield  {title} {\bibinfo {title} {{Noise in multiterminal
  diffusive conductors: Universality, nonlocality, and exchange effects}},\
  }\href@noop {} {\bibfield  {journal} {\bibinfo  {journal} {Phys. Rev. B}\
  }\textbf {\bibinfo {volume} {59}},\ \bibinfo {pages} {54} (\bibinfo {year}
  {1999})}\BibitemShut {NoStop}%
\bibitem [{\citenamefont {Tan}\ \emph {et~al.}(2018)\citenamefont {Tan},
  \citenamefont {Elo}, \citenamefont {Puska}, \citenamefont {Sarkar},
  \citenamefont {L{\"{a}}hteenm{\"{a}}ki}, \citenamefont {Duerr}, \citenamefont
  {Gould}, \citenamefont {Molenkamp}, \citenamefont {Nagaev},\ and\
  \citenamefont {Hakonen}}]{Tan2018}%
  \BibitemOpen
  \bibfield  {author} {\bibinfo {author} {\bibfnamefont {Z.~B.}\ \bibnamefont
  {Tan}}, \bibinfo {author} {\bibfnamefont {T.}~\bibnamefont {Elo}}, \bibinfo
  {author} {\bibfnamefont {A.}~\bibnamefont {Puska}}, \bibinfo {author}
  {\bibfnamefont {J.}~\bibnamefont {Sarkar}}, \bibinfo {author} {\bibfnamefont
  {P.}~\bibnamefont {L{\"{a}}hteenm{\"{a}}ki}}, \bibinfo {author}
  {\bibfnamefont {F.}~\bibnamefont {Duerr}}, \bibinfo {author} {\bibfnamefont
  {C.}~\bibnamefont {Gould}}, \bibinfo {author} {\bibfnamefont {L.~W.}\
  \bibnamefont {Molenkamp}}, \bibinfo {author} {\bibfnamefont {K.~E.}\
  \bibnamefont {Nagaev}},\ and\ \bibinfo {author} {\bibfnamefont {P.~J.}\
  \bibnamefont {Hakonen}},\ }\bibfield  {title} {\bibinfo {title}
  {{Hanbury-Brown and Twiss exchange and non-equilibrium-induced correlations
  in disordered, four-terminal graphene-ribbon conductor}},\ }\href
  {https://doi.org/10.1038/s41598-018-32777-5} {\bibfield  {journal} {\bibinfo
  {journal} {Sci. Rep.}\ }\textbf {\bibinfo {volume} {8}},\ \bibinfo {pages}
  {14952} (\bibinfo {year} {2018})}\BibitemShut {NoStop}%
\bibitem [{\citenamefont {Katsnelson}(2006)}]{Katsnelson2006}%
  \BibitemOpen
  \bibfield  {author} {\bibinfo {author} {\bibfnamefont {M.~I.}\ \bibnamefont
  {Katsnelson}},\ }\bibfield  {title} {\bibinfo {title} {{Zitterbewegung,
  chirality, and minimal conductivity in graphene}},\ }\href
  {https://doi.org/10.1140/epjb/e2006-00203-1} {\bibfield  {journal} {\bibinfo
  {journal} {Eur. Phys. J. B}\ }\textbf {\bibinfo {volume} {51}},\ \bibinfo
  {pages} {157} (\bibinfo {year} {2006})}\BibitemShut {NoStop}%
\bibitem [{\citenamefont {Tworzyd{\l}o}\ \emph {et~al.}(2006)\citenamefont
  {Tworzyd{\l}o}, \citenamefont {Trauzettel}, \citenamefont {Titov},
  \citenamefont {Rycerz},\ and\ \citenamefont {Beenakker}}]{Tworzydo2006}%
  \BibitemOpen
  \bibfield  {author} {\bibinfo {author} {\bibfnamefont {J.}~\bibnamefont
  {Tworzyd{\l}o}}, \bibinfo {author} {\bibfnamefont {B.}~\bibnamefont
  {Trauzettel}}, \bibinfo {author} {\bibfnamefont {M.}~\bibnamefont {Titov}},
  \bibinfo {author} {\bibfnamefont {A.}~\bibnamefont {Rycerz}},\ and\ \bibinfo
  {author} {\bibfnamefont {C.~W.~J.}\ \bibnamefont {Beenakker}},\ }\bibfield
  {title} {\bibinfo {title} {{Sub-Poissonian Shot Noise in Graphene}},\ }\href
  {https://doi.org/10.1103/PhysRevLett.96.246802} {\bibfield  {journal}
  {\bibinfo  {journal} {Phys. Rev. Lett.}\ }\textbf {\bibinfo {volume} {96}},\
  \bibinfo {pages} {246802} (\bibinfo {year} {2006})}\BibitemShut {NoStop}%
\bibitem [{\citenamefont {San-Jose}\ \emph {et~al.}(2007)\citenamefont
  {San-Jose}, \citenamefont {Prada},\ and\ \citenamefont
  {Golubev}}]{San-Jose2007}%
  \BibitemOpen
  \bibfield  {author} {\bibinfo {author} {\bibfnamefont {P.}~\bibnamefont
  {San-Jose}}, \bibinfo {author} {\bibfnamefont {E.}~\bibnamefont {Prada}},\
  and\ \bibinfo {author} {\bibfnamefont {D.~S.}\ \bibnamefont {Golubev}},\
  }\bibfield  {title} {\bibinfo {title} {{Universal scaling of current
  fluctuations in disordered graphene}},\ }\href
  {https://doi.org/10.1103/PhysRevB.76.195445} {\bibfield  {journal} {\bibinfo
  {journal} {Phys. Rev. B}\ }\textbf {\bibinfo {volume} {76}},\ \bibinfo
  {pages} {195445} (\bibinfo {year} {2007})}\BibitemShut {NoStop}%
\bibitem [{\citenamefont {Lewenkopf}\ \emph {et~al.}(2008)\citenamefont
  {Lewenkopf}, \citenamefont {Mucciolo},\ and\ \citenamefont {{Castro
  Neto}}}]{Lewenkopf2008}%
  \BibitemOpen
  \bibfield  {author} {\bibinfo {author} {\bibfnamefont {C.~H.}\ \bibnamefont
  {Lewenkopf}}, \bibinfo {author} {\bibfnamefont {E.~R.}\ \bibnamefont
  {Mucciolo}},\ and\ \bibinfo {author} {\bibfnamefont {A.~H.}\ \bibnamefont
  {{Castro Neto}}},\ }\bibfield  {title} {\bibinfo {title} {{Numerical studies
  of conductivity and Fano factor in disordered graphene}},\ }\href
  {https://doi.org/10.1103/PhysRevB.77.081410} {\bibfield  {journal} {\bibinfo
  {journal} {Phys. Rev. B}\ }\textbf {\bibinfo {volume} {77}},\ \bibinfo
  {pages} {081410} (\bibinfo {year} {2008})}\BibitemShut {NoStop}%
\bibitem [{\citenamefont {Laakso}\ and\ \citenamefont
  {Heikkil{\"{a}}}(2008)}]{Laakso2008}%
  \BibitemOpen
  \bibfield  {author} {\bibinfo {author} {\bibfnamefont {M.~A.}\ \bibnamefont
  {Laakso}}\ and\ \bibinfo {author} {\bibfnamefont {T.~T.}\ \bibnamefont
  {Heikkil{\"{a}}}},\ }\bibfield  {title} {\bibinfo {title} {{Charge transport
  in ballistic multiprobe graphene structures}},\ }\href
  {https://doi.org/10.1103/PhysRevB.78.205420} {\bibfield  {journal} {\bibinfo
  {journal} {Phys. Rev. B}\ }\textbf {\bibinfo {volume} {78}},\ \bibinfo
  {pages} {205420} (\bibinfo {year} {2008})}\BibitemShut {NoStop}%
\bibitem [{\citenamefont {DiCarlo}\ \emph {et~al.}(2008)\citenamefont
  {DiCarlo}, \citenamefont {Williams}, \citenamefont {Zhang}, \citenamefont
  {McClure},\ and\ \citenamefont {Marcus}}]{DiCarlo2008}%
  \BibitemOpen
  \bibfield  {author} {\bibinfo {author} {\bibfnamefont {L.}~\bibnamefont
  {DiCarlo}}, \bibinfo {author} {\bibfnamefont {J.~R.}\ \bibnamefont
  {Williams}}, \bibinfo {author} {\bibfnamefont {Y.}~\bibnamefont {Zhang}},
  \bibinfo {author} {\bibfnamefont {D.~T.}\ \bibnamefont {McClure}},\ and\
  \bibinfo {author} {\bibfnamefont {C.~M.}\ \bibnamefont {Marcus}},\ }\bibfield
   {title} {\bibinfo {title} {{Shot Noise in Graphene}},\ }\href
  {https://doi.org/10.1103/PhysRevLett.100.156801} {\bibfield  {journal}
  {\bibinfo  {journal} {Phys. Rev. Lett.}\ }\textbf {\bibinfo {volume} {100}},\
  \bibinfo {pages} {156801} (\bibinfo {year} {2008})}\BibitemShut {NoStop}%
\bibitem [{\citenamefont {Danneau}\ \emph
  {et~al.}(2008{\natexlab{a}})\citenamefont {Danneau}, \citenamefont {Wu},
  \citenamefont {Craciun}, \citenamefont {Russo}, \citenamefont {Tomi},
  \citenamefont {Salmilehto}, \citenamefont {Morpurgo},\ and\ \citenamefont
  {Hakonen}}]{Danneau2008}%
  \BibitemOpen
  \bibfield  {author} {\bibinfo {author} {\bibfnamefont {R.}~\bibnamefont
  {Danneau}}, \bibinfo {author} {\bibfnamefont {F.}~\bibnamefont {Wu}},
  \bibinfo {author} {\bibfnamefont {M.~F.}\ \bibnamefont {Craciun}}, \bibinfo
  {author} {\bibfnamefont {S.}~\bibnamefont {Russo}}, \bibinfo {author}
  {\bibfnamefont {M.~Y.}\ \bibnamefont {Tomi}}, \bibinfo {author}
  {\bibfnamefont {J.}~\bibnamefont {Salmilehto}}, \bibinfo {author}
  {\bibfnamefont {A.~F.}\ \bibnamefont {Morpurgo}},\ and\ \bibinfo {author}
  {\bibfnamefont {P.~J.}\ \bibnamefont {Hakonen}},\ }\bibfield  {title}
  {\bibinfo {title} {{Shot Noise in Ballistic Graphene}},\ }\href
  {https://doi.org/10.1103/PhysRevLett.100.196802} {\bibfield  {journal}
  {\bibinfo  {journal} {Phys. Rev. Lett.}\ }\textbf {\bibinfo {volume} {100}},\
  \bibinfo {pages} {196802} (\bibinfo {year} {2008}{\natexlab{a}})}\BibitemShut
  {NoStop}%
\bibitem [{\citenamefont {Danneau}\ \emph {et~al.}(2009)\citenamefont
  {Danneau}, \citenamefont {Wu}, \citenamefont {Craciun}, \citenamefont
  {Russo}, \citenamefont {Tomi}, \citenamefont {Salmilehto}, \citenamefont
  {Morpurgo},\ and\ \citenamefont {Hakonen}}]{Danneau2009}%
  \BibitemOpen
  \bibfield  {author} {\bibinfo {author} {\bibfnamefont {R.}~\bibnamefont
  {Danneau}}, \bibinfo {author} {\bibfnamefont {F.}~\bibnamefont {Wu}},
  \bibinfo {author} {\bibfnamefont {M.~F.}\ \bibnamefont {Craciun}}, \bibinfo
  {author} {\bibfnamefont {S.}~\bibnamefont {Russo}}, \bibinfo {author}
  {\bibfnamefont {M.~Y.}\ \bibnamefont {Tomi}}, \bibinfo {author}
  {\bibfnamefont {J.}~\bibnamefont {Salmilehto}}, \bibinfo {author}
  {\bibfnamefont {A.~F.}\ \bibnamefont {Morpurgo}},\ and\ \bibinfo {author}
  {\bibfnamefont {P.~J.}\ \bibnamefont {Hakonen}},\ }\bibfield  {title}
  {\bibinfo {title} {{Shot noise measurements in graphene}},\ }\href
  {https://doi.org/10.1016/j.ssc.2009.02.046} {\bibfield  {journal} {\bibinfo
  {journal} {Solid State Commun.}\ }\textbf {\bibinfo {volume} {149}},\
  \bibinfo {pages} {1050} (\bibinfo {year} {2009})}\BibitemShut {NoStop}%
\bibitem [{\citenamefont {Danneau}\ \emph {et~al.}(2010)\citenamefont
  {Danneau}, \citenamefont {Wu}, \citenamefont {Tomi}, \citenamefont
  {Oostinga}, \citenamefont {Morpurgo},\ and\ \citenamefont
  {Hakonen}}]{Danneau2010}%
  \BibitemOpen
  \bibfield  {author} {\bibinfo {author} {\bibfnamefont {R.}~\bibnamefont
  {Danneau}}, \bibinfo {author} {\bibfnamefont {F.}~\bibnamefont {Wu}},
  \bibinfo {author} {\bibfnamefont {M.~Y.}\ \bibnamefont {Tomi}}, \bibinfo
  {author} {\bibfnamefont {J.~B.}\ \bibnamefont {Oostinga}}, \bibinfo {author}
  {\bibfnamefont {a.~F.}\ \bibnamefont {Morpurgo}},\ and\ \bibinfo {author}
  {\bibfnamefont {P.~J.}\ \bibnamefont {Hakonen}},\ }\bibfield  {title}
  {\bibinfo {title} {{Shot noise suppression and hopping conduction in graphene
  nanoribbons}},\ }\href {https://doi.org/10.1103/PhysRevB.82.161405}
  {\bibfield  {journal} {\bibinfo  {journal} {Phys. Rev. B}\ }\textbf {\bibinfo
  {volume} {82}},\ \bibinfo {pages} {161405} (\bibinfo {year}
  {2010})}\BibitemShut {NoStop}%
\bibitem [{\citenamefont {Schottky}(1918)}]{Schottky1918}%
  \BibitemOpen
  \bibfield  {author} {\bibinfo {author} {\bibfnamefont {W.}~\bibnamefont
  {Schottky}},\ }\bibfield  {title} {\bibinfo {title} {{{\"{U}}ber spontane
  Stromschwankungen in verschiedenen Elektrizit{\"{a}}tsleitern}},\ }\href@noop
  {} {\bibfield  {journal} {\bibinfo  {journal} {Ann. Phys.}\ }\textbf
  {\bibinfo {volume} {57}},\ \bibinfo {pages} {541} (\bibinfo {year}
  {1918})}\BibitemShut {NoStop}%
\bibitem [{\citenamefont {Khlus}(1987)}]{Khlus1987}%
  \BibitemOpen
  \bibfield  {author} {\bibinfo {author} {\bibfnamefont {V.~A.}\ \bibnamefont
  {Khlus}},\ }\bibfield  {title} {\bibinfo {title} {{Current and voltage
  fluctuations in microjunctions between normal metals and superconductors}},\
  }\href
  {http://www.jetp.ac.ru/cgi-bin/dn/e{\_}066{\_}06{\_}1243.pdf{\%}5Cnhttp://jetp.ac.ru/cgi-bin/e/index/e/66/6/p1243?a=list}
  {\bibfield  {journal} {\bibinfo  {journal} {Sov. Phys JETP}\ }\textbf
  {\bibinfo {volume} {66}},\ \bibinfo {pages} {1243} (\bibinfo {year}
  {1987})}\BibitemShut {NoStop}%
\bibitem [{\citenamefont {Landauer}(1989)}]{Landauer1989}%
  \BibitemOpen
  \bibfield  {author} {\bibinfo {author} {\bibfnamefont {R.}~\bibnamefont
  {Landauer}},\ }\bibfield  {title} {\bibinfo {title} {{Johnson-Nyquist noise
  derived from quantum mechanical transmission}},\ }\href
  {https://doi.org/10.1016/0167-2789(89)90197-8} {\bibfield  {journal}
  {\bibinfo  {journal} {Phys. D}\ }\textbf {\bibinfo {volume} {38}},\ \bibinfo
  {pages} {226} (\bibinfo {year} {1989})}\BibitemShut {NoStop}%
\bibitem [{\citenamefont {Lesovik}(1989)}]{Lesovik1989}%
  \BibitemOpen
  \bibfield  {author} {\bibinfo {author} {\bibfnamefont {G.~B.}\ \bibnamefont
  {Lesovik}},\ }\bibfield  {title} {\bibinfo {title} {{Excess quantum noise in
  2D ballistic point contacts}},\ }\href@noop {} {\bibfield  {journal}
  {\bibinfo  {journal} {JETP Lett.}\ }\textbf {\bibinfo {volume} {49}},\
  \bibinfo {pages} {592} (\bibinfo {year} {1989})}\BibitemShut {NoStop}%
\bibitem [{\citenamefont {Yurke}\ and\ \citenamefont
  {Kochanski}(1990)}]{Yurke1990}%
  \BibitemOpen
  \bibfield  {author} {\bibinfo {author} {\bibfnamefont {B.}~\bibnamefont
  {Yurke}}\ and\ \bibinfo {author} {\bibfnamefont {G.~P.}\ \bibnamefont
  {Kochanski}},\ }\bibfield  {title} {\bibinfo {title} {{Momentum noise in
  vacuum tunneling transducers}},\ }\href
  {https://doi.org/10.1103/PhysRevB.41.8184} {\bibfield  {journal} {\bibinfo
  {journal} {Phys. Rev. B}\ }\textbf {\bibinfo {volume} {41}},\ \bibinfo
  {pages} {8184} (\bibinfo {year} {1990})}\BibitemShut {NoStop}%
\bibitem [{\citenamefont {B{\"{u}}ttiker}(1990)}]{Buttiker1990}%
  \BibitemOpen
  \bibfield  {author} {\bibinfo {author} {\bibfnamefont {M.}~\bibnamefont
  {B{\"{u}}ttiker}},\ }\bibfield  {title} {\bibinfo {title} {{Scattering theory
  of thermal and excess noise in open conductors}},\ }\href
  {https://doi.org/10.1103/PhysRevLett.65.2901} {\bibfield  {journal} {\bibinfo
   {journal} {Phys. Rev. Lett.}\ }\textbf {\bibinfo {volume} {65}},\ \bibinfo
  {pages} {2901} (\bibinfo {year} {1990})}\BibitemShut {NoStop}%
\bibitem [{\citenamefont {Beenakker}\ and\ \citenamefont
  {B{\"{u}}ttiker}(1992)}]{Beenakker1992}%
  \BibitemOpen
  \bibfield  {author} {\bibinfo {author} {\bibfnamefont {C.~W.~J.}\
  \bibnamefont {Beenakker}}\ and\ \bibinfo {author} {\bibfnamefont
  {M.}~\bibnamefont {B{\"{u}}ttiker}},\ }\bibfield  {title} {\bibinfo {title}
  {{Suppression of shot noise in metallic diffusive conductors}},\ }\href
  {https://doi.org/10.1103/PhysRevB.46.1889} {\bibfield  {journal} {\bibinfo
  {journal} {Phys. Rev. B}\ }\textbf {\bibinfo {volume} {46}},\ \bibinfo
  {pages} {1889} (\bibinfo {year} {1992})}\BibitemShut {NoStop}%
\bibitem [{\citenamefont {Nagaev}(1992)}]{Nagaev1992}%
  \BibitemOpen
  \bibfield  {author} {\bibinfo {author} {\bibfnamefont {K.}~\bibnamefont
  {Nagaev}},\ }\bibfield  {title} {\bibinfo {title} {{On the shot noise in
  dirty metal contacts}},\ }\href
  {https://doi.org/10.1016/0375-9601(92)90814-3} {\bibfield  {journal}
  {\bibinfo  {journal} {Phys. Lett. A}\ }\textbf {\bibinfo {volume} {169}},\
  \bibinfo {pages} {103} (\bibinfo {year} {1992})}\BibitemShut {NoStop}%
\bibitem [{\citenamefont {Fisher}\ and\ \citenamefont
  {Lee}(1981)}]{Fisher1981}%
  \BibitemOpen
  \bibfield  {author} {\bibinfo {author} {\bibfnamefont {D.~S.}\ \bibnamefont
  {Fisher}}\ and\ \bibinfo {author} {\bibfnamefont {P.~A.}\ \bibnamefont
  {Lee}},\ }\bibfield  {title} {\bibinfo {title} {{Relation between
  conductivity and transmission matrix}},\ }\href
  {https://doi.org/10.1103/PhysRevB.23.6851} {\bibfield  {journal} {\bibinfo
  {journal} {Phys. Rev. B}\ }\textbf {\bibinfo {volume} {23}},\ \bibinfo
  {pages} {6851} (\bibinfo {year} {1981})}\BibitemShut {NoStop}%
\bibitem [{\citenamefont {Aleiner}\ \emph {et~al.}(2002)\citenamefont
  {Aleiner}, \citenamefont {Brouwer},\ and\ \citenamefont
  {Glazman}}]{Aleiner2002}%
  \BibitemOpen
  \bibfield  {author} {\bibinfo {author} {\bibfnamefont {I.}~\bibnamefont
  {Aleiner}}, \bibinfo {author} {\bibfnamefont {P.}~\bibnamefont {Brouwer}},\
  and\ \bibinfo {author} {\bibfnamefont {L.}~\bibnamefont {Glazman}},\
  }\bibfield  {title} {\bibinfo {title} {{Quantum effects in Coulomb
  blockade}},\ }\href {https://doi.org/10.1016/S0370-1573(01)00063-1}
  {\bibfield  {journal} {\bibinfo  {journal} {Phys. Rep.}\ }\textbf {\bibinfo
  {volume} {358}},\ \bibinfo {pages} {309} (\bibinfo {year}
  {2002})}\BibitemShut {NoStop}%
\bibitem [{\citenamefont {Sukhorukov}\ and\ \citenamefont
  {Loss}(1998)}]{Sukhorukov1998}%
  \BibitemOpen
  \bibfield  {author} {\bibinfo {author} {\bibfnamefont {E.~V.}\ \bibnamefont
  {Sukhorukov}}\ and\ \bibinfo {author} {\bibfnamefont {D.}~\bibnamefont
  {Loss}},\ }\bibfield  {title} {\bibinfo {title} {{Universality of Shot Noise
  in Multiterminal Diffusive Conductors}},\ }\href
  {https://doi.org/10.1103/PhysRevLett.80.4959} {\bibfield  {journal} {\bibinfo
   {journal} {Phys. Rev. Lett.}\ }\textbf {\bibinfo {volume} {80}},\ \bibinfo
  {pages} {4959} (\bibinfo {year} {1998})}\BibitemShut {NoStop}%
\bibitem [{\citenamefont {Voutilainen}\ \emph {et~al.}(2011)\citenamefont
  {Voutilainen}, \citenamefont {Fay}, \citenamefont {H{\"{a}}kkinen},
  \citenamefont {Viljas}, \citenamefont {Heikkil{\"{a}}},\ and\ \citenamefont
  {Hakonen}}]{voutilainen2010}%
  \BibitemOpen
  \bibfield  {author} {\bibinfo {author} {\bibfnamefont {J.}~\bibnamefont
  {Voutilainen}}, \bibinfo {author} {\bibfnamefont {A.}~\bibnamefont {Fay}},
  \bibinfo {author} {\bibfnamefont {P.}~\bibnamefont {H{\"{a}}kkinen}},
  \bibinfo {author} {\bibfnamefont {J.~K.}\ \bibnamefont {Viljas}}, \bibinfo
  {author} {\bibfnamefont {T.~T.}\ \bibnamefont {Heikkil{\"{a}}}},\ and\
  \bibinfo {author} {\bibfnamefont {P.~J.}\ \bibnamefont {Hakonen}},\
  }\bibfield  {title} {\bibinfo {title} {{Energy relaxation in graphene and its
  measurement with supercurrent}},\ }\href
  {https://doi.org/10.1103/PhysRevB.84.045419} {\bibfield  {journal} {\bibinfo
  {journal} {Phys. Rev. B}\ }\textbf {\bibinfo {volume} {84}},\ \bibinfo
  {pages} {045419} (\bibinfo {year} {2011})}\BibitemShut {NoStop}%
\bibitem [{\citenamefont {Terr{\'{e}}s}\ \emph {et~al.}(2016)\citenamefont
  {Terr{\'{e}}s}, \citenamefont {Chizhova}, \citenamefont {Libisch},
  \citenamefont {Peiro}, \citenamefont {J{\"{o}}rger}, \citenamefont {Engels},
  \citenamefont {Girschik}, \citenamefont {Watanabe}, \citenamefont
  {Taniguchi}, \citenamefont {Rotkin}, \citenamefont {Burgd{\"{o}}rfer},\ and\
  \citenamefont {Stampfer}}]{Terres2016}%
  \BibitemOpen
  \bibfield  {author} {\bibinfo {author} {\bibfnamefont {B.}~\bibnamefont
  {Terr{\'{e}}s}}, \bibinfo {author} {\bibfnamefont {L.~A.}\ \bibnamefont
  {Chizhova}}, \bibinfo {author} {\bibfnamefont {F.}~\bibnamefont {Libisch}},
  \bibinfo {author} {\bibfnamefont {J.}~\bibnamefont {Peiro}}, \bibinfo
  {author} {\bibfnamefont {D.}~\bibnamefont {J{\"{o}}rger}}, \bibinfo {author}
  {\bibfnamefont {S.}~\bibnamefont {Engels}}, \bibinfo {author} {\bibfnamefont
  {A.}~\bibnamefont {Girschik}}, \bibinfo {author} {\bibfnamefont
  {K.}~\bibnamefont {Watanabe}}, \bibinfo {author} {\bibfnamefont
  {T.}~\bibnamefont {Taniguchi}}, \bibinfo {author} {\bibfnamefont {S.~V.}\
  \bibnamefont {Rotkin}}, \bibinfo {author} {\bibfnamefont {J.}~\bibnamefont
  {Burgd{\"{o}}rfer}},\ and\ \bibinfo {author} {\bibfnamefont {C.}~\bibnamefont
  {Stampfer}},\ }\bibfield  {title} {\bibinfo {title} {{Size quantization of
  Dirac fermions in graphene constrictions}},\ }\href
  {https://doi.org/10.1038/ncomms11528} {\bibfield  {journal} {\bibinfo
  {journal} {Nat. Commun.}\ }\textbf {\bibinfo {volume} {7}},\ \bibinfo {pages}
  {11528} (\bibinfo {year} {2016})}\BibitemShut {NoStop}%
\bibitem [{\citenamefont {{Krishna Kumar}}\ \emph {et~al.}(2017)\citenamefont
  {{Krishna Kumar}}, \citenamefont {Bandurin}, \citenamefont {Pellegrino},
  \citenamefont {Cao}, \citenamefont {Principi}, \citenamefont {Guo},
  \citenamefont {Auton}, \citenamefont {{Ben Shalom}}, \citenamefont
  {Ponomarenko}, \citenamefont {Falkovich}, \citenamefont {Watanabe},
  \citenamefont {Taniguchi}, \citenamefont {Grigorieva}, \citenamefont
  {Levitov}, \citenamefont {Polini},\ and\ \citenamefont
  {Geim}}]{KrishnaKumar2017}%
  \BibitemOpen
  \bibfield  {author} {\bibinfo {author} {\bibfnamefont {R.}~\bibnamefont
  {{Krishna Kumar}}}, \bibinfo {author} {\bibfnamefont {D.~A.}\ \bibnamefont
  {Bandurin}}, \bibinfo {author} {\bibfnamefont {F.~M.}\ \bibnamefont
  {Pellegrino}}, \bibinfo {author} {\bibfnamefont {Y.}~\bibnamefont {Cao}},
  \bibinfo {author} {\bibfnamefont {A.}~\bibnamefont {Principi}}, \bibinfo
  {author} {\bibfnamefont {H.}~\bibnamefont {Guo}}, \bibinfo {author}
  {\bibfnamefont {G.~H.}\ \bibnamefont {Auton}}, \bibinfo {author}
  {\bibfnamefont {M.}~\bibnamefont {{Ben Shalom}}}, \bibinfo {author}
  {\bibfnamefont {L.~A.}\ \bibnamefont {Ponomarenko}}, \bibinfo {author}
  {\bibfnamefont {G.}~\bibnamefont {Falkovich}}, \bibinfo {author}
  {\bibfnamefont {K.}~\bibnamefont {Watanabe}}, \bibinfo {author}
  {\bibfnamefont {T.}~\bibnamefont {Taniguchi}}, \bibinfo {author}
  {\bibfnamefont {I.~V.}\ \bibnamefont {Grigorieva}}, \bibinfo {author}
  {\bibfnamefont {L.~S.}\ \bibnamefont {Levitov}}, \bibinfo {author}
  {\bibfnamefont {M.}~\bibnamefont {Polini}},\ and\ \bibinfo {author}
  {\bibfnamefont {A.~K.}\ \bibnamefont {Geim}},\ }\bibfield  {title} {\bibinfo
  {title} {{Superballistic flow of viscous electron fluid through graphene
  constrictions}},\ }\href {https://doi.org/10.1038/nphys4240} {\bibfield
  {journal} {\bibinfo  {journal} {Nat. Phys.}\ }\textbf {\bibinfo {volume}
  {13}},\ \bibinfo {pages} {1182} (\bibinfo {year} {2017})}\BibitemShut
  {NoStop}%
\bibitem [{\citenamefont {Cleric{\`{o}}}\ \emph {et~al.}(2018)\citenamefont
  {Cleric{\`{o}}}, \citenamefont {Delgado-Notario}, \citenamefont
  {Saiz-Bret{\'{i}}n}, \citenamefont {{Hern{\'{a}}ndez Fuentevilla}},
  \citenamefont {Malyshev}, \citenamefont {Lejarreta}, \citenamefont {Diez},\
  and\ \citenamefont {Dom{\'{i}}nguez-Adame}}]{Clerico2018}%
  \BibitemOpen
  \bibfield  {author} {\bibinfo {author} {\bibfnamefont {V.}~\bibnamefont
  {Cleric{\`{o}}}}, \bibinfo {author} {\bibfnamefont {J.~A.}\ \bibnamefont
  {Delgado-Notario}}, \bibinfo {author} {\bibfnamefont {M.}~\bibnamefont
  {Saiz-Bret{\'{i}}n}}, \bibinfo {author} {\bibfnamefont {C.}~\bibnamefont
  {{Hern{\'{a}}ndez Fuentevilla}}}, \bibinfo {author} {\bibfnamefont {A.~V.}\
  \bibnamefont {Malyshev}}, \bibinfo {author} {\bibfnamefont {J.~D.}\
  \bibnamefont {Lejarreta}}, \bibinfo {author} {\bibfnamefont {E.}~\bibnamefont
  {Diez}},\ and\ \bibinfo {author} {\bibfnamefont {F.}~\bibnamefont
  {Dom{\'{i}}nguez-Adame}},\ }\bibfield  {title} {\bibinfo {title} {{Quantized
  Electron Transport Through Graphene Nanoconstrictions}},\ }\href
  {https://doi.org/10.1002/pssa.201701065} {\bibfield  {journal} {\bibinfo
  {journal} {Phys. Status Solidi}\ }\textbf {\bibinfo {volume} {215}},\
  \bibinfo {pages} {1701065} (\bibinfo {year} {2018})}\BibitemShut {NoStop}%
\bibitem [{\citenamefont {{Das Sarma}}\ \emph {et~al.}(2011)\citenamefont {{Das
  Sarma}}, \citenamefont {Adam}, \citenamefont {Hwang},\ and\ \citenamefont
  {Rossi}}]{DasSarma2011}%
  \BibitemOpen
  \bibfield  {author} {\bibinfo {author} {\bibfnamefont {S.}~\bibnamefont {{Das
  Sarma}}}, \bibinfo {author} {\bibfnamefont {S.}~\bibnamefont {Adam}},
  \bibinfo {author} {\bibfnamefont {E.~H.}\ \bibnamefont {Hwang}},\ and\
  \bibinfo {author} {\bibfnamefont {E.}~\bibnamefont {Rossi}},\ }\bibfield
  {title} {\bibinfo {title} {{Electronic transport in two-dimensional
  graphene}},\ }\href {https://doi.org/10.1103/RevModPhys.83.407} {\bibfield
  {journal} {\bibinfo  {journal} {Rev. Mod. Phys.}\ }\textbf {\bibinfo {volume}
  {83}},\ \bibinfo {pages} {407} (\bibinfo {year} {2011})}\BibitemShut
  {NoStop}%
\bibitem [{Note1()}]{Note1}%
  \BibitemOpen
  \bibinfo {note} {Comsol Multiphysics was used for the
  calculation.}\BibitemShut {Stop}%
\bibitem [{\citenamefont {Nieminen}\ \emph {et~al.}(2016)\citenamefont
  {Nieminen}, \citenamefont {L{\"{a}}hteenm{\"{a}}ki}, \citenamefont {Tan},
  \citenamefont {Cox},\ and\ \citenamefont {Hakonen}}]{Nieminen2016}%
  \BibitemOpen
  \bibfield  {author} {\bibinfo {author} {\bibfnamefont {T.}~\bibnamefont
  {Nieminen}}, \bibinfo {author} {\bibfnamefont {P.}~\bibnamefont
  {L{\"{a}}hteenm{\"{a}}ki}}, \bibinfo {author} {\bibfnamefont
  {Z.}~\bibnamefont {Tan}}, \bibinfo {author} {\bibfnamefont {D.}~\bibnamefont
  {Cox}},\ and\ \bibinfo {author} {\bibfnamefont {P.~J.}\ \bibnamefont
  {Hakonen}},\ }\bibfield  {title} {\bibinfo {title} {{Low-noise correlation
  measurements based on software-defined-radio receivers and cooled microwave
  amplifiers}},\ }\href {https://doi.org/10.1063/1.4966971} {\bibfield
  {journal} {\bibinfo  {journal} {Rev. Sci. Instrum.}\ }\textbf {\bibinfo
  {volume} {87}},\ \bibinfo {pages} {114706} (\bibinfo {year}
  {2016})}\BibitemShut {NoStop}%
\bibitem [{\citenamefont {Danneau}\ \emph
  {et~al.}(2008{\natexlab{b}})\citenamefont {Danneau}, \citenamefont {Wu},
  \citenamefont {Craciun}, \citenamefont {Russo}, \citenamefont {Tomi},
  \citenamefont {Salmilehto}, \citenamefont {Morpurgo},\ and\ \citenamefont
  {Hakonen}}]{Danneau2008a}%
  \BibitemOpen
  \bibfield  {author} {\bibinfo {author} {\bibfnamefont {R.}~\bibnamefont
  {Danneau}}, \bibinfo {author} {\bibfnamefont {F.}~\bibnamefont {Wu}},
  \bibinfo {author} {\bibfnamefont {M.~F.}\ \bibnamefont {Craciun}}, \bibinfo
  {author} {\bibfnamefont {S.}~\bibnamefont {Russo}}, \bibinfo {author}
  {\bibfnamefont {M.~Y.}\ \bibnamefont {Tomi}}, \bibinfo {author}
  {\bibfnamefont {J.}~\bibnamefont {Salmilehto}}, \bibinfo {author}
  {\bibfnamefont {A.~F.}\ \bibnamefont {Morpurgo}},\ and\ \bibinfo {author}
  {\bibfnamefont {P.~J.}\ \bibnamefont {Hakonen}},\ }\bibfield  {title}
  {\bibinfo {title} {{Evanescent Wave Transport and Shot Noise in Graphene:
  Ballistic Regime and Effect of Disorder}},\ }\href
  {https://doi.org/10.1007/s10909-008-9837-z} {\bibfield  {journal} {\bibinfo
  {journal} {J. Low Temp. Phys.}\ }\textbf {\bibinfo {volume} {153}},\ \bibinfo
  {pages} {374} (\bibinfo {year} {2008}{\natexlab{b}})}\BibitemShut {NoStop}%
\bibitem [{\citenamefont {van Langen}\ and\ \citenamefont
  {B{\"{u}}ttiker}(1997)}]{VanLangen1997}%
  \BibitemOpen
  \bibfield  {author} {\bibinfo {author} {\bibfnamefont {S.~A.}\ \bibnamefont
  {van Langen}}\ and\ \bibinfo {author} {\bibfnamefont {M.}~\bibnamefont
  {B{\"{u}}ttiker}},\ }\bibfield  {title} {\bibinfo {title}
  {{Quantum-statistical current correlations in multilead chaotic cavities}},\
  }\href {https://doi.org/10.1103/PhysRevB.56.R1680} {\bibfield  {journal}
  {\bibinfo  {journal} {Phys. Rev. B}\ }\textbf {\bibinfo {volume} {56}},\
  \bibinfo {pages} {R1680} (\bibinfo {year} {1997})}\BibitemShut {NoStop}%
\bibitem [{\citenamefont {Laitinen}\ \emph {et~al.}(2016)\citenamefont
  {Laitinen}, \citenamefont {Paraoanu}, \citenamefont {Oksanen}, \citenamefont
  {Craciun}, \citenamefont {Russo}, \citenamefont {Sonin},\ and\ \citenamefont
  {Hakonen}}]{Laitinen2016}%
  \BibitemOpen
  \bibfield  {author} {\bibinfo {author} {\bibfnamefont {A.}~\bibnamefont
  {Laitinen}}, \bibinfo {author} {\bibfnamefont {G.~S.}\ \bibnamefont
  {Paraoanu}}, \bibinfo {author} {\bibfnamefont {M.}~\bibnamefont {Oksanen}},
  \bibinfo {author} {\bibfnamefont {M.~F.}\ \bibnamefont {Craciun}}, \bibinfo
  {author} {\bibfnamefont {S.}~\bibnamefont {Russo}}, \bibinfo {author}
  {\bibfnamefont {E.}~\bibnamefont {Sonin}},\ and\ \bibinfo {author}
  {\bibfnamefont {P.}~\bibnamefont {Hakonen}},\ }\bibfield  {title} {\bibinfo
  {title} {{Contact doping, Klein tunneling, and asymmetry of shot noise in
  suspended graphene}},\ }\href {https://doi.org/10.1103/PhysRevB.93.115413}
  {\bibfield  {journal} {\bibinfo  {journal} {Phys. Rev. B}\ }\textbf {\bibinfo
  {volume} {93}},\ \bibinfo {pages} {1} (\bibinfo {year} {2016})}\BibitemShut
  {NoStop}%
\bibitem [{\citenamefont {Nagaev}(1995)}]{Nagaev1995}%
  \BibitemOpen
  \bibfield  {author} {\bibinfo {author} {\bibfnamefont {K.~E.}\ \bibnamefont
  {Nagaev}},\ }\bibfield  {title} {\bibinfo {title} {{Influence of
  electron-electron scattering on shot noise in diffusive contacts}},\ }\href
  {https://doi.org/10.1103/physrevb.52.4740} {\bibfield  {journal} {\bibinfo
  {journal} {Phys. Rev. B}\ }\textbf {\bibinfo {volume} {52}},\ \bibinfo
  {pages} {4740} (\bibinfo {year} {1995})}\BibitemShut {NoStop}%
\bibitem [{\citenamefont {Chen}\ \emph {et~al.}(2008)\citenamefont {Chen},
  \citenamefont {Jang}, \citenamefont {Xiao}, \citenamefont {Ishigami},\ and\
  \citenamefont {Fuhrer}}]{Chen2008}%
  \BibitemOpen
  \bibfield  {author} {\bibinfo {author} {\bibfnamefont {J.-H.}\ \bibnamefont
  {Chen}}, \bibinfo {author} {\bibfnamefont {C.}~\bibnamefont {Jang}}, \bibinfo
  {author} {\bibfnamefont {S.}~\bibnamefont {Xiao}}, \bibinfo {author}
  {\bibfnamefont {M.}~\bibnamefont {Ishigami}},\ and\ \bibinfo {author}
  {\bibfnamefont {M.~S.}\ \bibnamefont {Fuhrer}},\ }\bibfield  {title}
  {\bibinfo {title} {{Intrinsic and extrinsic performance limits of graphene
  devices on SiO2}},\ }\href {https://doi.org/10.1038/nnano.2008.58} {\bibfield
   {journal} {\bibinfo  {journal} {Nat. Nanotechnol.}\ }\textbf {\bibinfo
  {volume} {3}},\ \bibinfo {pages} {206} (\bibinfo {year} {2008})}\BibitemShut
  {NoStop}%
\bibitem [{\citenamefont {Fratini}\ and\ \citenamefont
  {Guinea}(2008)}]{PolarModes}%
  \BibitemOpen
  \bibfield  {author} {\bibinfo {author} {\bibfnamefont {S.}~\bibnamefont
  {Fratini}}\ and\ \bibinfo {author} {\bibfnamefont {F.}~\bibnamefont
  {Guinea}},\ }\bibfield  {title} {\bibinfo {title} {{Substrate-limited
  electron dynamics in graphene}},\ }\href
  {https://doi.org/10.1103/PhysRevB.77.195415} {\bibfield  {journal} {\bibinfo
  {journal} {Phys. Rev. B}\ }\textbf {\bibinfo {volume} {77}},\ \bibinfo
  {pages} {195415} (\bibinfo {year} {2008})}\BibitemShut {NoStop}%
\end{thebibliography}%

\end{document}